%% file: ms.tex
\documentclass[prd,reprint,showpacs,superscriptaddress,bibnotes]{revtex4-1}
\usepackage{natbib}
\usepackage{graphics}
\usepackage{multirow}
\usepackage{amsbsy}
\usepackage{mathrsfs}
\usepackage{footmisc}
\usepackage{hyperref}
\hypersetup{
  colorlinks = true,
  urlcolor = black,
  linkcolor=black,
  citecolor=black
}
\input{defs}

\begin{document}

\title{Detection of \textit{B}-Mode Polarization at Degree Angular Scales by \biceptwo}

\author{P.~A.~R.~\surname{Ade}}
\affiliation{School of Physics and Astronomy, Cardiff University, Cardiff, CF24 3AA, United Kingdom}
\author{R.~W.~Aikin}
\affiliation{Department of Physics, California Institute of Technology, Pasadena, California 91125, USA}
\author{D.~Barkats}
\affiliation{Joint ALMA Observatory, Vitacura, Santiago, Chile}
\author{S.~J.~Benton}
\affiliation{Department of Physics, University of Toronto, Toronto, Ontario, M5S 1A7, Canada}
\author{C.~A.~Bischoff}
\affiliation{Harvard-Smithsonian Center for Astrophysics, 60 Garden Street MS 42, Cambridge, Massachusetts 02138, USA}
\author{J.~J.~Bock}
\affiliation{Department of Physics, California Institute of Technology, Pasadena, California 91125, USA}
\affiliation{Jet Propulsion Laboratory, Pasadena, California 91109, USA}
\author{J.~A.~Brevik}
\affiliation{Department of Physics, California Institute of Technology, Pasadena, California 91125, USA}
\author{I.~Buder}
\affiliation{Harvard-Smithsonian Center for Astrophysics, 60 Garden Street MS 42, Cambridge, Massachusetts 02138, USA}
\author{E.~Bullock}
\affiliation{Minnesota Institute for Astrophysics, University of Minnesota, Minneapolis, Minnesota 55455, USA}
\author{C.~D.~Dowell}
\affiliation{Jet Propulsion Laboratory, Pasadena, California 91109, USA}
\author{L.~Duband}
\affiliation{Service des Basses Temp\'{e}ratures, Commissariat \`{a} l'Energie Atomique, 38054 Grenoble, France}
\author{J.~P.~Filippini}
\affiliation{Department of Physics, California Institute of Technology, Pasadena, California 91125, USA}
\author{S.~Fliescher}
\affiliation{Department of Physics, University of Minnesota, Minneapolis, Minnesota 55455, USA}
\author{S.~R.~Golwala}
\affiliation{Department of Physics, California Institute of Technology, Pasadena, California 91125, USA}
\author{M.~Halpern}
\affiliation{Department of Physics and Astronomy, University of British Columbia, Vancouver, British Columbia, V6T 1Z1, Canada}
\author{M.~Hasselfield}
\affiliation{Department of Physics and Astronomy, University of British Columbia, Vancouver, British Columbia, V6T 1Z1, Canada}
\author{S.~R.~Hildebrandt}
\affiliation{Department of Physics, California Institute of Technology, Pasadena, California 91125, USA}
\affiliation{Jet Propulsion Laboratory, Pasadena, California 91109, USA}
\author{G.~C.~Hilton}
\affiliation{National Institute of Standards and Technology, Boulder, Colorado 80305, USA}
\author{V.~V.~Hristov}
\affiliation{Department of Physics, California Institute of Technology, Pasadena, California 91125, USA}
\author{K.~D.~Irwin}
\affiliation{Department of Physics, Stanford University, Stanford, California 94305, USA}
\affiliation{Kavli Institute for Particle Astrophysics and Cosmology, SLAC National Accelerator Laboratory, 2575 Sand Hill Rd, Menlo Park, California 94025, USA}
\affiliation{National Institute of Standards and Technology, Boulder, Colorado 80305, USA}
\author{K.~S.~Karkare}
\affiliation{Harvard-Smithsonian Center for Astrophysics, 60 Garden Street MS 42, Cambridge, Massachusetts 02138, USA}
\author{J.~P.~Kaufman}
\affiliation{Department of Physics, University of California at San Diego, La Jolla, California 92093, USA}
\author{B.~G.~Keating}
\affiliation{Department of Physics, University of California at San Diego, La Jolla, California 92093, USA}
\author{S.~A.~Kernasovskiy}
\affiliation{Department of Physics, Stanford University, Stanford, California 94305, USA}
\author{J.~M.~Kovac}
\email{jmkovac@cfa.harvard.edu}
\affiliation{Harvard-Smithsonian Center for Astrophysics, 60 Garden Street MS 42, Cambridge, Massachusetts 02138, USA}
\author{C.~L.~Kuo}
\affiliation{Department of Physics, Stanford University, Stanford, California 94305, USA}
\affiliation{Kavli Institute for Particle Astrophysics and Cosmology, SLAC National Accelerator Laboratory, 2575 Sand Hill Rd, Menlo Park, California 94025, USA}
\author{E.~M.~Leitch}
\affiliation{University of Chicago, Chicago, Illinois 60637, USA}
\author{M.~Lueker}
\affiliation{Department of Physics, California Institute of Technology, Pasadena, California 91125, USA}
\author{P.~Mason}
\affiliation{Department of Physics, California Institute of Technology, Pasadena, California 91125, USA}
\author{C.~B.~Netterfield}
\affiliation{Department of Physics, University of Toronto, Toronto, Ontario, M5S 1A7, Canada}
\affiliation{Canadian Institute for Advanced Research, Toronto, Ontario, M5G 1Z8, Canada}
\author{H.~T.~Nguyen}
\affiliation{Jet Propulsion Laboratory, Pasadena, California 91109, USA}
\author{R.~O'Brient}
\affiliation{Jet Propulsion Laboratory, Pasadena, California 91109, USA}
\author{R.~W.~Ogburn~IV}
\affiliation{Department of Physics, Stanford University, Stanford, California 94305, USA}
\affiliation{Kavli Institute for Particle Astrophysics and Cosmology, SLAC National Accelerator Laboratory, 2575 Sand Hill Rd, Menlo Park, California 94025, USA}
\author{A.~Orlando}
\affiliation{Department of Physics, University of California at San Diego, La Jolla, California 92093, USA}
\author{C.~Pryke}
\email{pryke@physics.umn.edu}
\affiliation{Department of Physics, University of Minnesota, Minneapolis, Minnesota 55455, USA}
\affiliation{Minnesota Institute for Astrophysics, University of Minnesota, Minneapolis, Minnesota 55455, USA}
\author{C.~D.~Reintsema}
\affiliation{National Institute of Standards and Technology, Boulder, Colorado 80305, USA}
\author{S.~Richter}
\affiliation{Harvard-Smithsonian Center for Astrophysics, 60 Garden Street MS 42, Cambridge, Massachusetts 02138, USA}
\author{R.~Schwarz}
\affiliation{Department of Physics, University of Minnesota, Minneapolis, Minnesota 55455, USA}
\author{C.~D.~Sheehy}
\affiliation{Department of Physics, University of Minnesota, Minneapolis, Minnesota 55455, USA}
\affiliation{University of Chicago, Chicago, Illinois 60637, USA}
\author{Z.~K.~Staniszewski}
\affiliation{Department of Physics, California Institute of Technology, Pasadena, California 91125, USA}
\affiliation{Jet Propulsion Laboratory, Pasadena, California 91109, USA}
\author{R.~V.~Sudiwala}
\affiliation{School of Physics and Astronomy, Cardiff University, Cardiff, CF24 3AA, United Kingdom}
\author{G.~P.~Teply}
\affiliation{Department of Physics, California Institute of Technology, Pasadena, California 91125, USA}
\author{J.~E.~Tolan}
\affiliation{Department of Physics, Stanford University, Stanford, California 94305, USA}
\author{A.~D.~Turner}
\affiliation{Jet Propulsion Laboratory, Pasadena, California 91109, USA}
\author{A.~G.~Vieregg}
\affiliation{Harvard-Smithsonian Center for Astrophysics, 60 Garden Street MS 42, Cambridge, Massachusetts 02138, USA}
\affiliation{University of Chicago, Chicago, Illinois 60637, USA}
\author{C.~L.~Wong}
\affiliation{Harvard-Smithsonian Center for Astrophysics, 60 Garden Street MS 42, Cambridge, Massachusetts 02138, USA}
\author{K.~W.~Yoon}
\affiliation{Department of Physics, Stanford University, Stanford, California 94305, USA}
\affiliation{Kavli Institute for Particle Astrophysics and Cosmology, SLAC National Accelerator Laboratory, 2575 Sand Hill Rd, Menlo Park, California 94025, USA}
\collaboration{\biceptwo\ Collaboration}

\date[Published~]{19 June 2014}

\begin{abstract}
We report results from the \biceptwo\ experiment,
a cosmic microwave background (CMB) polarimeter specifically designed
to search for the signal of inflationary gravitational waves
in the \bmode\ power spectrum around $\ell \sim 80$.
The telescope comprised a 26~cm aperture all-cold refracting
optical system equipped with a focal plane
of 512 antenna coupled transition edge sensor 150 GHz bolometers each
with temperature sensitivity of $\approx300$~\ukrts.
\biceptwo\ observed from the South Pole for three seasons from 2010 to 2012.
A low-foreground region of sky with an effective area of 380 square deg
was observed to a depth of 87~nK deg in Stokes $Q$ and $U$.
In this paper we describe the observations, data reduction, maps,
simulations, and results.
We find an excess of \bmode\ power over the base lensed-\lcdm\ expectation
in the range $30<\ell<150$, inconsistent with the null hypothesis
at a significance of $>5\sigma$.
Through jackknife tests and simulations based on detailed
calibration measurements we show that systematic contamination
is much smaller than the observed excess.
Cross correlating against \wmap\ 23~GHz maps we find that Galactic synchrotron
makes a negligible contribution to the observed signal.
We also examine a number of available models of polarized dust emission
and find that at their default parameter values they predict power 
$\sim(5-10)\times$ smaller than the observed excess signal
(with no significant cross-correlation with our maps).
However, these models are not sufficiently constrained by external public data
to exclude the possibility of dust emission bright enough
to explain the entire excess signal.
Cross correlating \biceptwo\ against 100~GHz maps from the \bicepone\ experiment,
the excess signal is confirmed with $3\sigma$ significance and
its spectral index is found to be consistent with that of the CMB,
disfavoring dust at $1.7\sigma$.
The observed \bmode\ power spectrum is well fit by a
lensed-\lcdm\ + tensor theoretical model with tensor-to-scalar ratio
$r=0.20^{+0.07}_{-0.05}$,
with $r=0$ disfavored at $7.0\sigma$.
Accounting for the contribution of foreground dust will shift
this value downward by an amount which will be better constrained
with upcoming datasets.
\end{abstract}

\doi{10.1103/PhysRevLett.112.241101}

\maketitle

\section{Introduction}

\setcounter{footnote}{1}

The discovery of the cosmic microwave background (CMB)
by Penzias and Wilson \cite{penzias65} confirmed the hot big bang paradigm
and established the CMB as a central tool for the study of cosmology.
In recent years, observations of its temperature anisotropies have helped establish
and refine the ``standard'' cosmological model now known as \lcdm,
under which our universe is understood to be spatially flat,
dominated by cold dark matter, and with a cosmological constant
($\Lambda$) driving accelerated expansion at late times.
CMB temperature measurements have now reached
remarkable precision over angular scales ranging
from the whole sky to arcmin resolution,
producing results in striking concordance with predictions of
\lcdm\ and constraining its key parameters to sub-percent precision
(e.g.,~\cite{bennett13,hinshaw13,story13,hou14,sievers13,das13,planckXV,planckXVI}).

Inflationary cosmology extends the standard model by
postulating an early period of nearly exponential expansion
which sets the initial conditions for the subsequent hot big bang.
It was proposed and developed in the early 1980s to resolve
mysteries for which the standard model offered no solution,
including the flatness, horizon, smoothness, entropy, and monopole problems
(\cite{brout78,starobinsky80,kazanas80,sato81,guth81,linde82,linde83,albrecht82};
see \cite{planckXXII} for a review).
Inflation also explains the Universe's primordial perturbations
as originating in quantum fluctuations stretched by this
exponential expansion
\cite{mukhanov81,hawking82,guth82,starobinsky82,bardeen83,mukhanov85}, 
and thus to be correlated on superhorizon scales.
The simplest models further predict these perturbations to be highly adiabatic and Gaussian
and nearly scale-invariant (though typically slightly stronger on larger scales).
These qualities, though not necessarily unique to the inflationary paradigm,
have all been confirmed by subsequent observations
(e.g.,~\cite{spergel97,peiris03}, and references above).
Although highly successful, the inflationary paradigm represents
a vast extrapolation from well-tested regimes in physics.
It invokes quantum effects in highly curved spacetime at
energies near $10^{16}$~GeV and timescales less than $10^{-32}$~s.
A definitive test of this paradigm would be of fundamental importance.

Gravitational waves generated by inflation 
have the potential to provide such a definitive test.
Inflation predicts that the quantization of the gravitational field
coupled to exponential expansion produces a primordial background
of stochastic gravitational waves with a characteristic spectral shape
(\cite{grishchuk75a,starobinsky79,rubakov82,fabbri83,abbott84};
also see \cite{ashoorioon12,krauss13}).
Though unlikely to be directly detectable in modern instruments, these gravitational 
waves would have imprinted a unique signature upon the CMB.
Gravitational waves induce local 
quadrupole anisotropies in the radiation field within the last-scattering 
surface, inducing polarization in the scattered light~\cite{polnarev85}.
This polarization pattern will include a ``curl'' or \bmode\ component at degree 
angular scales that cannot be generated primordially by density perturbations.  
The amplitude of this signal depends upon the tensor-to-scalar ratio~$r$, which 
itself is a function of the energy scale of inflation.
The detection of \bmode\ polarization of the CMB at large angular scales would
provide a unique confirmation of inflation and a probe of its energy 
scale~\cite{seljak97b,kamionkowski97,seljak97a}.

The CMB is polarized with an amplitude of a few $\mu$K,
dominated by the ``gradient'' or \emode\ pattern that is generated by density 
perturbations at last scattering.
These $E$ modes peak at angular scales
of $\sim0.2\deg$, corresponding to angular multipole $\ell\approx 1000$.
They were first detected by the \dasi\
experiment~\cite{kovac02}.
Since then multiple experiments have refined measurements of the
$EE$ power spectrum, including
\capmap~\cite{barkats04,bischoff08}, 
\cbi~\cite{readhead04,sievers07}, 
\boom03~\cite{montroy06}, 
\wmap~\cite{page07,bennett13}, 
\maxipol~\cite{wu07}, 
\QUAD~\cite{pryke09,brown09}, %
\bicepone~\cite{chiang10,barkats14}, 
and \quiet~\cite{quiet11,quiet12}. 

Gravitational lensing of the CMB's light by 
large scale structure at relatively late times
produces small deflections of the primordial pattern,
converting a small portion of \emode\ power into $B$ modes.
The lensing \bmode\ spectrum is similar to a smoothed version
of the \emode\ spectrum but a factor $\sim100$ lower in power,
and hence also rises toward subdegree scales and peaks around $\ell\approx 1000$.
The inflationary gravitational wave (IGW) $B$ mode, however, is predicted to peak at
multipole $\ell\approx 80$ and this creates an opportunity to
search for it around this scale where it is quite distinct from
the lensing effect. (This is the so-called ``recombination bump.''
There is another opportunity to search for the IGW signal
at $\ell<10$ in the ``reionization bump,'' but this
requires observations over a substantial fraction of the full
sky.)

A large number of current CMB experimental efforts now target
\bmode\ polarization.  Evidence for lensing \bmode\ polarization
at subdegree scales has already been detected by two experiments in the
past year, first by the \spt\ telescope~\cite{hanson13}
and more recently by 
\polarbear ~\cite{polarbear6645,polarbear6646,polarbear14}.
The search for inflationary $B$ modes at larger scales will also
be a goal of these experiments, as well as other ongoing experimental
efforts in the U.S. that include the 
\abs~\cite{hileman09},
\actpol~\cite{niemack10}, and 
\class~\cite{class12} ground-based telescopes and the 
\ebex~\cite{reichborn10}, 
\spider~\cite{fraisse13}, and 
\piper~\cite{kogut12}
balloon experiments, 
each employing a variety of complementary strategies.
It is also a major science goal of the ESA \planck\ satellite mission.

The \bicep /\keckarray\ series of experiments have been specifically designed to search
for primordial \bmode\ polarization on degree angular scales
by making very deep maps of relatively small patches of sky
from the South Pole.
The \bicepone\ instrument initiated this series~\cite{Keating03},
observing from 2006 to 2008.
Its initial results were described in Takahashi \textit{et al.}~\cite{takahashi10}
and Chiang \textit{et al.}~\cite{chiang10} (hereafter T10 and C10),
and final results were recently reported in Barkats \textit{et al.}~\cite{barkats14} (hereafter B14)
yielding a limit of $r<0.70$ at 95\% confidence.

In this paper we report results from \biceptwo---a successor to
\bicepone\ which differed principally in the focal
plane where a very large increase in the detector count resulted
in more than an order of magnitude improvement in mapping speed.
The observation field and strategy were largely unchanged,
as were the telescope mount, observation site, etc.
Using all three seasons of data taken with \biceptwo\ (2010--2012) we
detect \bmode\ power in the multipole range $30<\ell<150$, finding this
power to have a strong excess inconsistent with lensed \lcdm\
at $>5\sigma$ significance.

The structure of this paper is as follows.
In Secs.~\ref{sec:instrument} and~\ref{sec:obs} we briefly review the
\biceptwo\ instrument, observations, and low-level
data reduction deferring details to a related
paper~\cite{b2instpap14} (hereafter the
Instrument Paper).
In Sec.~\ref{sec:mapmaking} we describe our map-making 
procedure and present signal and signal-differenced
``jackknife'' $T$, $Q$, and $U$ maps which have unprecedented sensitivity.
This section introduces ``deprojection'' of modes
potentially contaminated through beam systematics,
which is an important new technique.
In Sec.~\ref{sec:sims} we describe our detailed
time stream-level simulations of signal and pseudosimulations of
noise.
In Sec.~\ref{sec:mapstospectra} we describe calculation of
the power spectra, including matrix-based \bmode\ purification.
In Sec.~\ref{sec:powspecres} we present the signal and
jackknife power spectrum results for $TE$, $EE$, $BB$, $TB$, and
$EB$.
In Sec.~\ref{sec:sysuncer} we discuss and summarize the
many studies we have conducted probing for actual and
potential sources of systematic contamination, and argue
that residual contamination is much smaller
than the detected \bmode\ signal.
Full details are deferred to a related paper
\cite{b2systpap14} (hereafter the Systematics Paper).
In Sec.~\ref{sec:foregrounds} we investigate foreground
projections and constraints based on external data and conclude
that it is implausible that the \bmode\ signal which we
see is dominated by synchrotron, and that the present
data disfavor domination by dust or any other known foreground source.
In Sec.~\ref{sec:cross_spectra} we take cross spectra of
the \biceptwo\ maps with those from \bicepone\
(as presented in B14) and find that the spectral signature
of the signal is consistent with the CMB.
Finally in Sec.~\ref{sec:parcons} we calculate some
simple, largely phenomenological, parameter constraints,
and conclude in Sec.~\ref{sec:conc}.

\section{The \biceptwo\ Instrument}
\label{sec:instrument}

\biceptwo\ was similar to \bicepone\ (see T10)
reusing the same telescope mount and installation at the South Pole.
Like \bicepone\ the optical system was a simple 26~cm aperture all-cold
refractor housed entirely in a liquid helium cooled cryostat.
The main differences from \bicepone\ were the use of a focal plane
array of planar antenna-coupled devices~\cite{bock03} with voltage-biased transition-edge
sensor (TES) detectors~\cite{irwin95} and a multiplexed superconducting quantum interference
device (SQUID) readout.  \biceptwo\ observed at 150~GHz only.
A very brief review of the instrument follows---for more
details please refer to the Instrument Paper.

\subsection{Optics}

The optics were adapted from the original \bicepone\ design~\cite{Keating03}.
Light entered the cryostat through a polypropylene foam
window, passed through polytetrafluoroethylene filters cooled to 100~K and 40~K,
and then through polyethylene objective and eyepiece lenses
cooled to 4~K.
A 26.4~cm diameter aperture stop was placed at the objective lens
and an additional nylon filter was placed on the sky side of the
eyepiece lens.
All the lenses and filters were antireflection coated and
the interior of the optics tube was lined with microwave
absorber.
The optics were designed to be telecentric (flat focal plane)
and the resulting beams had full width at half maximum of $\approx0.5\deg$.
An absorptive fore baffle was mounted on the front of the telescope
which was designed to prevent radiation from boresight angles
greater than $\sim20\deg$ entering the telescope.
The telescope was located inside a large stationary
reflective ground screen.

\subsection{Focal plane}

The \biceptwo\ focal plane employed monolithic arrays
of antenna coupled TES detectors designed and fabricated
at Caltech and JPL.
Each pixel was composed of two interleaved 12$\times$12 arrays of
orthogonal slot antennas feeding beam-forming (phased-array) summing trees.
The output of each summing tree was a microstrip which
passed through a band-defining filter and deposited
its power on a thermally isolated island.
Changes in the power incident on this island were detected using a
transition edge sensor (TES).
There was an 8$\times$8 array of pixels on each tile,
and four such tiles were combined to form the complete
focal plane unit.
There were thus, in principle, 256 dual-polarization pixels in the
focal plane for a total of 512 detectors,
each with temperature sensitivity of $\approx 300$~\ukrts.
(Six pixels were deliberately disconnected
between antenna and TES sensor to provide diagnostic ``dark'' channels.)
The focal plane was cooled to 270~mK by a closed
cycle three-stage sorption refrigerator.

\subsection{Detector readout and data acquisition system}

The TES detectors were read out through time-division
SQUID multiplexing chips provided by NIST.
A single readout channel was connected in rapid succession
to 32 detectors, reducing wiring and heat load requirements.
These SQUID systems were biased and read out by a multi channel 
electronics (MCE) crate 
external to the cryostat (provided by UBC).
The sample rate stored to disk was 20~Hz.
The housekeeping and readout electronics were connected
to a set of Linux-based computers running a control
system called GCP, which has been used by many 
recent ground-based CMB experiments~\cite{story2012}.

\subsection{Telescope mount}

The receiver cryostat was mounted on a three-axis mount
able to move in azimuth and elevation and to rotate the entire
telescope about its boresight.
Hereafter, we refer to the line-of-sight rotation angle as the
``deck'' angle.
The window of the telescope looked out through an opening
in a flexible environmental seal such that the cryostat,
mount, and electronics were all located in a room temperature
laboratory environment.

\section{Observations and Low-Level Data Reduction}
\label{sec:obs}

\subsection{Observations}
\label{sec:obsobs}

\biceptwo\ observed on a three day schedule locked
to sidereal time.
As in \bicepone, the basic unit of observation was a $\approx$ 50~minute
``scan set'' during which the telescope scanned back-and-forth
53 times at $2.8\deg$~s$^{-1}$ in azimuth in a smooth turnaround triangle wave pattern,
with a throw of $\approx 60\deg$, at fixed elevation.
We refer to each of the 106 motions across the field
(either left- or right-going) as a ``half scan''.
We do not use the turnaround portions of the scans in this analysis.

\biceptwo\ observed the same CMB field as \bicepone---a 
low foreground region centered at RA 0h, Dec.\ $-57.5\deg$.
At the South Pole, the elevation angle is simply the negative
of declination and azimuth maps to RA shifting by
$15\deg$ per hour.
The scan speed on the sky was thus $\approx 1.5\deg$~s$^{-1}$
mapping multipole $\ell=100$ into the timestream at $\approx 0.4$~Hz.
At the end of each scan set the elevation was stepped by $0.25\deg$
and the azimuth angle updated to recenter on RA 0h.
The scans thus ``slide'' with respect to the sky during each
scan set by $\approx 12.5\deg$ allowing us to subtract
a scan fixed ``template'' from the time stream while
leaving degree-scale sky structure only slightly
attenuated (see Secs.~\ref{sec:filt} and~\ref{sec:supfac}).

A total of 21 elevation offsets were used between
Dec.\ of $-55\deg$ and $-60\deg$.
Note that since the field of view of the 
focal plane---$\sim 20\deg$---is much larger
in Dec., and somewhat larger in RA, than the region scanned
by the boresight the final coadded map is naturally apodized.
After a complete three-day schedule the instrument
was rotated to a new deck angle, the refrigerator was recycled,
and the process repeated.
See the Instrument Paper for more details of the observation
strategy.

The control system ran CMB observation schedules relentlessly
between early 2010 and late 2012 collecting
over 17~000 scansets of data ($\approx590$~days).
(There were some breaks for beam mapping and other calibrations
during the austral summers.)
The raw data were transferred off site daily via satellite, allowing
rigorous quality monitoring and ongoing analysis development.
The analysis presented in this paper uses all of the CMB data
taken by \biceptwo.

\subsection{Analysis pipeline}

The analysis pipeline used in this paper is written in the
\textsc{matlab} language and was originally developed for the \QUAD\
experiment~\cite{pryke09}.
It was then adapted to \bicepone\ data and became the secondary,
and then primary, analysis pipeline for the C10 and B14 papers,
respectively.
For \biceptwo\ it has seen substantial further development
including the addition of a sophisticated automatic data selection
framework, full deprojection of beam systematics, and
a map-based \bmode\ purification operation; these
enhancements are detailed below.

\subsection{Transfer function correction and deglitching}

Starting from the raw time streams, the first step of the pipeline
is to deconvolve the temporal response of the instrument.
The TES detectors themselves have a very fast and uniform
response at all frequencies of interest.
To correct for the effect of the digital low-pass filtering,
which was applied to the data before it was down sampled for
recording to disk, we apply an FIR deconvolution operation in the
time domain (which also reapplies a zero-delay
low pass filter).
Glitches and flux jumps in the SQUID readout are also corrected
and/or flagged
at this point---they are relatively rare in these data.
See the Instrument Paper for more details.

\subsection{Relative gain calibration}

At the beginning and end of each scan set an elevation
nod or ``elnod'' was performed.
The telescope was moved up-down-up or vice versa in
a roughly sinusoidal excursion in time, injecting
a signal proportional to the
atmospheric opacity gradient into the detector time streams.
In analysis, each elnod is regressed against the air-mass profile
through which it was looking to derive a relative
gain coefficient in SQUID feedback units per air mass.
The time stream for each scan set is then divided by its own elnod coefficient
and multiplied by the median over all good detectors.
This roughly equalizes the gain of each channel
and results in considerable cancellation of
atmospheric fluctuations when taking the difference of detector pairs,
thus making the data considerably easier to work with.
The relative gain as determined using the atmospheric
gradient is not necessarily the relative gain
which minimizes leakage of CMB temperature anisotropy
to polarization---see Sec.~\ref{sec:sysuncer}.
Absolute calibration is deferred until after the final
coadded map is made---see Sec.~\ref{sec:maps}.

\subsection{First round data cuts}

At this point in the data reduction, individual
channels are cut at per half-scan granularity.
Reasons for removal include glitches and flux jumps
in the channel in question, or its multiplex neighbors,
and synchronization problems in the data acquisition system.
\biceptwo\ data are very well behaved and over 90\% of the data pass
this stage.

\section{Map Making}
\label{sec:mapmaking}

\subsection{Time stream filtering}
\label{sec:filt}

In the next step the sum and difference of each detector
pair is taken, the pair sum being ultimately used to form
maps of temperature anisotropy, and the pair difference
to measure polarization.
Each half scan is then subjected to a third-order polynomial filtering.

Each half scan is constrained to have the same number
of time samples.
In addition to the polynomial filtering
we also perform a ``template'' subtraction
of any scan-synchronous component by averaging together the
corresponding points over a scan set and removing the result
from each half scan.
Forward and backward half scans are treated separately.

Within our simulation-based analysis framework
we are free to perform any arbitrary filtering of
the data which we choose.
Although any given filtering implies some loss of
sensitivity due to the removal and mixing of modes within
the map these effects are corrected as described in Secs.~\ref{sec:matpure}
and~\ref{sec:supfac}.
We defer discussion of the particular filtering choices made in this
analysis to Sec.~\ref{sec:sysuncer}.

\subsection{Pointing reconstruction}

The pointing trajectory of the telescope boresight
(i.e., the line-of-sight axis of rotation of the mount)
is determined using a mount pointing model calibrated using a star camera
as described in the Instrument Paper.
To convert time stream into maps it is then necessary to
know the pointing offset of each detector from this
direction.
To measure these we first make per
channel maps assuming approximate offsets, and then
regress these against the \wmap5 temperature map to determine corrections.
Comparing maps made from left-going and right-going scans
at each of the four deck angles, we estimate
that this procedure is accurate to better than $0.05\deg$ absolute pointing
uncertainty.
The beam positions relative to the boresight are averaged
over the scan directions and deck angles to produce
a single reconstruction for each detector
used in map making.

\subsection{Construction of deprojection time stream}
\label{sec:deprojtemplates}

The two halves of each detector pair would ideally
have identical angular response patterns (beams) on the sky.
If this is not the case, then leakage of temperature
anisotropy (pair sum) to polarization (pair difference)
will occur~\cite{shimon08}.
One can resample an external map of the temperature
sky and its derivatives to generate templates
of the leakage resulting from specific differential
beam effects.
In this analysis we smooth the \planck\ 143~GHz map~\footnote{
\url{http://irsa.ipac.caltech.edu/data/Planck/release_1/all-sky-maps/previews/HFI_SkyMap_143_2048_R1.10_nominal/index.html}} 
using the average measured beam function and resample following the procedure
described in Ref.~\cite{sheehy13} and the Systematics Paper.
Our standard procedure is to
calculate templates for the six modes which correspond
to differences of elliptical Gaussian beams.
In practice we do not normally use all six---see 
Secs.~\ref{sec:deproj} and~\ref{sec:sysuncer}.

\subsection{Binning into pair maps}

At this point we bin the pair-sum and pair-difference signals
into per-scan set, per-pair RA-Dec.\ pixel grids which we refer
to as ``pair maps.''
The pixels are $0.25\deg$ square at declination $-57.5\deg$.
The data from each scan set are weighted by
their inverse variance over the complete scan set (with
separate weights for pair sum and pair difference).
We note that while the pair-sum weights vary widely
due to variation in atmospheric $1/f$ noise, the pair-difference
weights are extremely stable over time---i.e., atmospheric
fluctuations are empirically shown to be highly unpolarized.
For pair difference a number of products of the time stream and
the sine and cosine of the polarization angle are recorded
to allow construction of $Q$ and $U$ maps as described in Sec.~\ref{sec:acctomaps}
below.
The deprojection templates are also binned into pair maps
in parallel with the pair-difference data.

We use per-pair detector polarization angles derived from a
dielectric sheet calibrator (as described in the Instrument Paper). 
(These derived angles are within $0.2^\circ$ rms of their design values,
well within the required accuracy.
However note that we later apply an overall rotation
to minimize the high $\ell$ $TB$ and $EB$ spectra---see Sec.~\ref{sec:polrot}.)
The measured polarization efficiency of our detectors is very high
($\approx 99$\%, see the Instrument Paper)---we perform a small correction
to convert temperature-based gains to polarization gains.

\subsection{Second round data cuts}

The per-scan set, per-pair maps are recorded on disk to allow
rapid recalculation of the coadded map while varying
the so-called ``second round'' cut parameters.
These include a variety of cuts on the bracketing
elnods, including goodness of fit to the atmospheric cosecant model
and stability in both absolute and pair-relative senses.
We also make some cuts based on the behavior of the data
themselves, including tests for skewness and noise stationarity.
Many of these cuts identify periods of exceptionally
bad weather and are redundant with one another.
We also apply ``channel cuts'' to remove a small
fraction of pairs---principally those with 
anomalous measured differential beam shapes.
In general \biceptwo\ data are very well behaved and
the final fraction of data retained is 63\%.
See the Instrument Paper for more details.

\subsection{Accumulation of pair maps to phase, and template regression}
\label{sec:deproj}

Once the second round cuts have been made we accumulate
the pair maps over each set of ten elevation steps
(hereafter referred to as a ``phase'').
The deprojection templates are also accumulated.
We then regress some of these binned templates against the
data---i.e., we effectively find the best fit value
for each nonideality, for each pair, within
each phase.
The templates scaled by the regression coefficients are
then subtracted from the data, entirely removing that
imperfection mode if present.
This operation also filters real signal and noise
due to chance correlation (and real $TE$-induced 
correlation in the case of signal).
This filtering is effectively just additional
time stream filtering like that already mentioned
in Sec.~\ref{sec:filt} and we calibrate and correct
for its effect in the same way (see Secs.~\ref{sec:matpure}
and~\ref{sec:supfac}).

The choice of deprojection time scale is a compromise---reducing 
it guards against systematic modes which
vary over short time scales (as relative gain errors might),
while covering more sky before regressing reduces the
filtering of real signal (the coefficient is fit to a greater number
of pixels).
In practice reduction of the filtering going from ten
elevation steps to twenty is found to be modest
and for this analysis we have deprojected modes on
a per-phase basis.

We also have the option to fix the coefficients
of any given mode at externally measured values, corresponding to a subtraction
of the systematic with no additional filtering of signal.
In this analysis we have deprojected differential
gain and pointing, and have subtracted the effects of differential 
ellipticity---we 
defer discussion of these particular choices to Sec.~\ref{sec:sysuncer}.

\subsection{Accumulation over phases and pairs}
\label{sec:acctomaps}

We next proceed to coadded maps accumulating
over phases and pairs.
Full coadds are produced as well as many ``jackknife'' splits---pairs 
of maps made from two subsets of the data which
might be hypothesized to contain different systematic
contaminations.
Some splits are strictly temporal (e.g., first half vs second
half of the observations), some are strictly pair selections
(e.g., inner vs outer part of each detector tile), and
some are both temporal and pairwise (e.g., the so-called tile and deck
jackknife)---see Sec.~\ref{sec:jacks} for details.

Once the accumulation over all 590~days and $\approx 200$~detector pairs
is done the accumulated quantities must be converted
to $T$, $Q$ and $U$ maps.
For $T$ this is as simple as dividing by the sum of the weights.
For $Q$ and $U$ we must perform a simple $2\times2$ matrix inversion
for each pixel.
This matrix is singular if a given pixel has been observed at
only a single value of the deck angle modulo $90\deg$.
In general for \biceptwo\ data we have angles
$68\deg$, $113\deg$, $248\deg$, and $293\deg$
as measured relative to the celestial meridian.

We perform absolute calibration by taking the cross spectrum
of the $T$ map with either the \planck\ 143~GHz map or the \wmap9 
$W$-band $T$ map as described in the Instrument Paper.
We adopt an absolute gain value intermediate to these two measurements
and assign calibration uncertainty of 1.3\% in the map to account for
the difference.

\subsection{Maps}
\label{sec:maps}

\begin{figure*}
\resizebox{\textwidth}{!}{\includegraphics{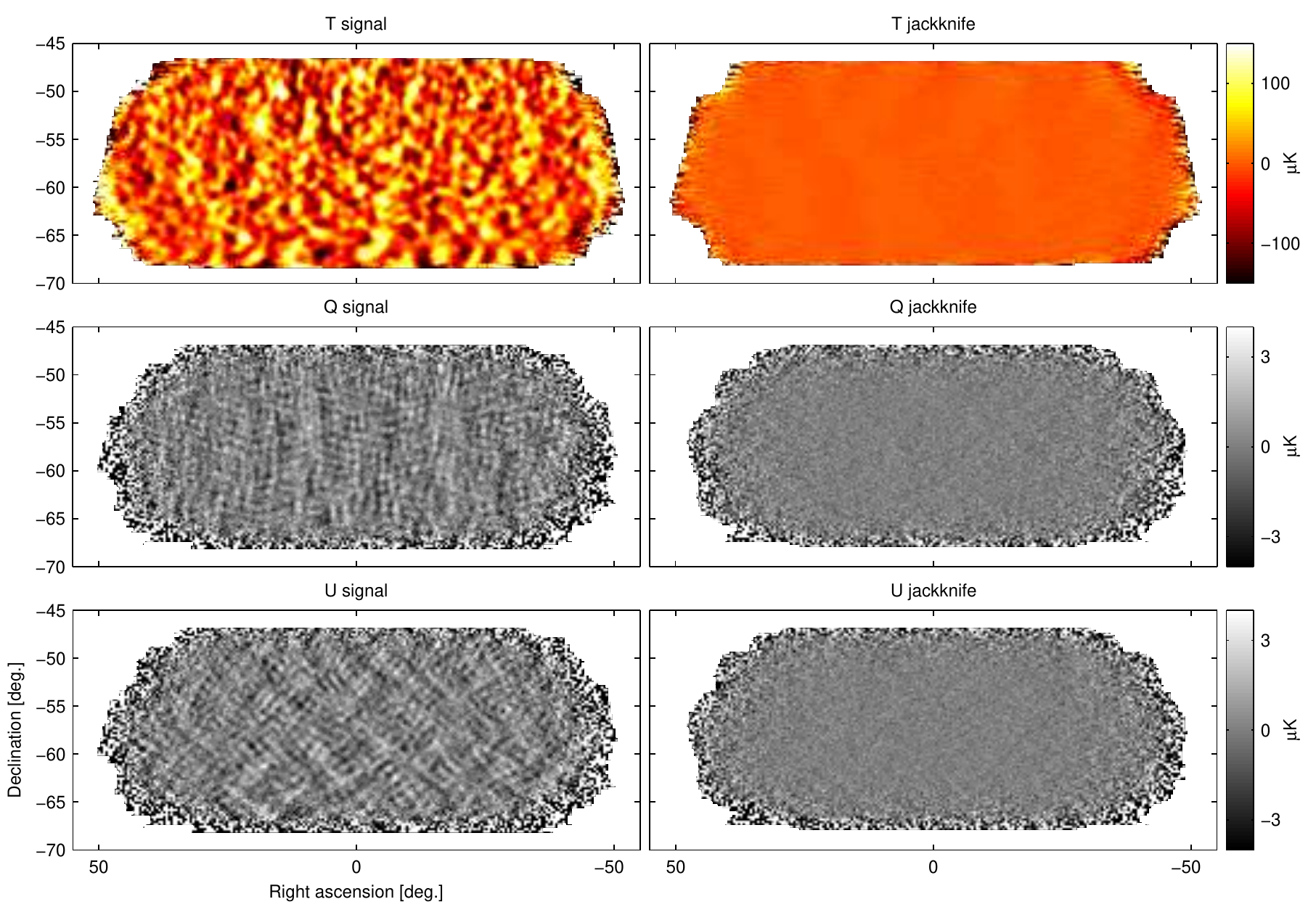}}
\caption{\biceptwo\ $T$, $Q$, $U$ maps.
The left column shows the basic signal maps with $0.25\deg$ pixelization
as output by the reduction pipeline.
The right column shows difference (jackknife) maps made with the
first and second halves of the data set.
No additional filtering other than that imposed by the instrument beam
(FWHM $0.5\deg$) has been done.
Note that the structure seen in the $Q$ and $U$ signal maps is as expected
for an \emode\ dominated sky.}
\label{fig:tqu_maps}
\end{figure*}

Figure~\ref{fig:tqu_maps} shows the \biceptwo\ $T$, $Q$, and
$U$ signal maps along with a sample set of difference (jackknife) maps.
The ``vertical-stripe-$Q$, diagonal-stripe-$U$''
pattern indicative of an \emode\ dominated sky is visible.
Note that these maps are filtered by the relatively
large beam of \biceptwo\ ($\approx 0.5\deg$ FWHM).
Comparison of the signal and jackknife maps shows
that the former are signal dominated---they are the deepest maps
of CMB polarization ever made at degree angular scales with an rms noise
level of 87~nK in (nominal) $1\deg \times 1\deg$ pixels.

\section{Simulations}
\label{sec:sims}

\subsection{Signal simulations}

As is common practice in this type of analysis, we account for the
filtering which our instrument and data reduction impose
on the underlying sky pattern through simulations~\cite{hivon02}.
Starting with input $T$, $Q$, and $U$ sky maps we smooth
using the average measured beam function and then resample along the pointing
trajectory of each detector at each time stream sample.
We have the option of perturbing to per-channel elliptical
Gaussian beam shapes
using the derivatives of the map (in a similar manner to the
construction of the deprojection templates described
in Sec.~\ref{sec:deprojtemplates} above).
However, for our standard simulations we include only differential pointing as
this is our leading order beam imperfection (see Sec.~\ref{sec:sysuncer}).

We perform three sets of signal-only simulations:
(i) simulations generated from unlensed \lcdm\ input spectra (hereafter ``unlensed \lcdm''),
(ii) simulations generated from those same input skies, explicitly lensed in
map space as described below (hereafter ``lensed \lcdm''),
and (iii) simulations containing only tensor $B$ modes with $r=0.2$ (and $n_t=0$).

\subsubsection{Constrained input maps}
\label{sec:inputmaps}

The observing matrix and purification operator described in Sec.~\ref{sec:matpure} 
are constructed for a specific assumed $T$ sky map.
Since its construction is computationally very expensive
it is preferable to constrain the input $T$ skies used for the
simulations to be the same rather than to recalculate the
operator for each simulation.

To construct constrained $Q$ and $U$ sky maps which respect
the known \lcdm\ $TE$ correlation we start from a map of the 
well-measured temperature anisotropy, specifically the \planck\ 
needlet internal linear combination (NILC) $T$ map~\footnote{
\url{http://irsa.ipac.caltech.edu/data/Planck/release_1/all-sky-maps/previews/COM_CompMap_CMB-nilc_2048_R1.20/index.html}}.
We calculate the $a_{\ell m}^T$ using the \synfast\ software from the \healpix~\footnote{
\url{http://healpix.sourceforge.net/}}
 package~\cite{healpix}, and then calculate sets
of $a_{\ell m}^E$ using

\begin{equation}\label{eqn:constrained}
a_{\ell m}^E=\frac{C_{\ell}^{TE}}{C_{\ell}^{TT}}a_{\ell m}^T+\sqrt{C_{\ell}^{EE}-(C_{\ell}^{TE})^2/C_{\ell}^{TT}}n_{\ell m}^{}
\end{equation}

\noindent where the $C_{\ell}$'s are \lcdm\ spectra from
\camb~\footnote{\url{http://camb.info/}} with cosmological
parameters taken from \planck~\cite{planckXVI}, and the $n_{\ell m}$ are
normally distributed complex random numbers.
For $C_{\ell}^{TT}$ we use a lensed-\lcdm\ spectrum since the
$a_{\ell m}^T$ from \planck\ NILC inherently contain lensing.
We have found the noise level in the \planck\
NILC maps for our region of observation and multipole range to be low enough
that it can be ignored.

Using the $a_{\ell m}^E$ we generate Nside~=~2048 maps using \synfast.
We substitute in the $a_{\ell m}^T$ from \planck\ 143~GHz so that the $T$ map 
more closely resembles the $T$ sky we expect to see with \biceptwo. 
(This is also the map that is used in Sec.~\ref{sec:deproj} 
to construct deprojection templates.)
Additionally, we add in noise to the $T$ map at the level
predicted by the noise covariance in the \planck\ 143~GHz map, which allows 
us to simulate any deprojection residual due to noise
in the \planck\ 143~GHz map.

\subsubsection{Lensing of input maps}
\label{sec:lensedmaps}

Lensing is added to the unlensed-\lcdm\ maps using the
\lenspix~\footnote{\url{http://cosmologist.info/lenspix/}} software~\cite{lewis05}.
We use this software to generate lensed versions
of the constrained CMB input $a_{\ell m}$'s described in Sec.~\ref{sec:inputmaps}.
Input to the lensing operation are deflection angle spectra that
are generated with \camb\ as part of the standard computation of \lcdm\ spectra.
The lensing operation is performed before the beam smoothing
is applied to form the final map products. We do not apply
lensing to the 143~GHz temperature $a_{\ell m}^T$ from \planck\ since these
inherently contain lensing. Our simulations hence approximate lensed
CMB maps ignoring the lensing correlation between $T$ and $E$.

\subsection{Noise pseudosimulations}
\label{sec:noisesims}

The previous \bicepone\ and \QUAD\ pipelines used a Fourier
based procedure
to make simulated noise time streams, maintaining
correlations between all channels~\cite{pryke09}.
For the increased channel count in \biceptwo\ this
is computationally very expensive, so we have switched to
an alternate procedure adapted from \spt~\cite{vanengelen12}.
We perform additional coadds of the real pair maps randomly
flipping the sign of each scanset.
The sign-flip sequences are constructed such that
the total weight of positively and negatively weighted maps is equal.
We have checked this technique against the older
technique, and against another technique which constructs
map noise covariance matrices, and have found them all to be
equivalent to the relevant level of accuracy.
[In the lowest two band powers a difference can be detected within
the available statistics between the sign-flip and traditional
noise generators, with the sign flip predicting $(10-15)$\%
higher noise power.
This is about one third of the fluctuation on the noise, 
and about 5\% of the apparent signal. The sign-flip and
matrix techniques agree to within the available statistics.
Since the sign-flip sequences are 17~000 scansets long the
resulting maps are effectively uncorrelated.
Separate sequences are used for each half of each temporal
jackknife.]
By default we use the sign-flipping technique
and refer to these realizations as ``noise pseudosimulations.''

We add the noise maps to the lensed-\lcdm\ realizations
to form signal plus noise simulations---hereafter
referred to as lensed-\lcdm+noise.

\section{From Maps to Power Spectra}
\label{sec:mapstospectra}

\subsection{Inversion to spectra}

The most basic power spectrum estimation
procedure which one can employ is to apply an apodization
windowing, Fourier transform, construct $E$ and $B$ from $Q$ and $U$,
square, and take the means in annuli
as estimates of the CMB band powers.
A good choice for the window may be the inverse of the
noise variance map (or a smoothed version thereof).
Employing this simple procedure on the unlensed-\lcdm\
simulations we find an unacceptable degree of $E$ to $B$
mixing.
While such mixing can be corrected for in the mean using
simulations, its fluctuation leads to a significant loss of
sensitivity.

There are several things which can cause $E$ to $B$ mixing:
(i) the ``sky cut'' implied by the apodization window
(the transformation from $Q$ and $U$ to $E$ and $B$ is
nonlocal so some of the modes around the edge of the map are ambiguous),
(ii) the time stream (and therefore map) filtering which we have imposed
in Secs.~\ref{sec:filt} and~\ref{sec:deproj}, and
(iii) the simple RA-Dec.\ map projection which we have chosen.

To correct for sky cut-induced mixing, improved estimators
have been suggested.
We first tried implementing the estimator suggested by Smith~\cite{smith06}
which takes Fourier transforms of products of the map
with various derivatives of the apodization window.
However, testing on the unlensed-\lcdm\ simulations revealed
only a modest improvement in performance since this estimator
does not correct mixing caused by filtering of the map.

\subsection{Matrix-based map purification}
\label{sec:matpure}

To overcome the $E$ to $B$ mixing described in the previous
subsection we have introduced an additional purification step after the $Q$ and $U$ maps are formed.
This step has to be performed in pixel space where the filtering takes place.
In parallel with the construction of the pair maps
and their accumulation we construct
pixel-pixel matrices which track how every true sky pixel
maps into the pixels of our final coadded map due to
the various filtering operations.
We take ``true sky pixel maps''
to be Nside~=~512 \healpix\ maps, whose pixel size ($\sim0.1^\circ$ on a side)
is smaller than our observed map pixels ($0.25^\circ$).
The act of simulating our various
filtering operations becomes a simple matrix multiplication:
\begin{equation}\label{eqn:obs}
\mathbf{\tilde m}=\mathbf{Rm}
\end{equation}
where $\mathbf{m}$ is a vector consisting of [$Q,U$] values for
each \healpix\ pixel and $\mathbf{\tilde m}$ is a [$Q,U$] vector
as observed by \biceptwo\ in the absence of noise.

Next, we ``observe'' an Nside~=~512 \healpix\
theoretical covariance matrix
(constructed following Appendix A of Ref.~\cite{tegmark01}),
$\mathbf{C}$, with $\mathbf{R}$:
\begin{equation}\label{eqn:obs_cov}
\mathbf{\tilde C}=\mathbf{RCR}^\mathrm{T}
\end{equation}

We form $\mathbf{\tilde{C}}$ for both \emode\ and
\bmode\ covariances.
These matrices provide the pixel-pixel covariance for $E$ modes
and $B$ modes in the same observed space as the real data.
However, the matrix $\mathbf{R}$ has made the two spaces nonorthogonal
and introduced ambiguous modes, i.e., modes in the observed space which
are superpositions of either $E$ modes or $B$ modes on the sky.

To isolate the pure $B$ modes we adapt the method described in Bunn \textit{et al.}~\cite{bunn03}.
We solve a generalized eigenvalue problem:
\begin{equation}
\label{eqn:gen_eigen}
(\mathbf{\tilde C_B}+\sigma^2 \mathbf{I})\mathbf{b}=\lambda_{\mathbf{b}}(\mathbf{\tilde C_E}+\sigma^2 \mathbf{I})\mathbf{b}
\end{equation}
where $\mathbf{b}$ is a [$Q,U$] eigenmode and $\sigma^2$ is 
a small number introduced to regularize the problem. 
By selecting modes corresponding to the largest
eigenvalues $\lambda_{\mathbf{b}} \gg 1$, we can find the \bmode\ subspace
of the observed sky which is orthogonal to $E$ modes and
ambiguous modes.
The covariance matrices are calculated using steeply reddened input spectra
($\sim 1/{\ell}^2$) so that the eigenmodes are separated in
angular scale, making it easy to select modes up to a cutoff
$\ell$ set by the instrument resolution.

The matrix purification operator is a sum of outer 
products of the selected eigenmodes; it projects
an input map onto this space of pure $B$ modes:
\begin{equation}
\mathbf{\Pi_b} = \sum_i \mathbf{b}_i \mathbf{b}_i^{\mathrm{T}}
\end{equation}
It can be applied to any simulated map vector ($\mathbf{\tilde{m}}$)
and returns a purified vector which contains only
signal coming from $B$ modes on the true sky:
\begin{equation}\label{eqn:proj}
\mathbf{\tilde m}_{\mathrm{pure}} = \mathbf{\Pi_b \tilde m}
\end{equation}
This method is superior to the other methods discussed above because it
removes the $E$-to-$B$ leakage resulting from 
the filterings and the sky-cut,
because $\mathbf{R}$ contains all of these steps.
After the purification, in the present analysis we
use the simple power spectrum estimation described in
the previous subsection, although in the future
we may switch to a fully matrix-based approach.

Testing this operator on the standard unlensed-\lcdm\ simulations
(which are constructed entirely independently) we empirically
determine that it is extremely effective, with residual
false $B$ modes corresponding to $r<10^{-4}$.
Testing the operator on the $r=0.2$ simulations we find
that it produces only a very modest increase in the sample
variance---i.e., the fraction of mixed (ambiguous) modes is
found to be small.

\subsection{Noise subtraction and filter and beam correction}
\label{sec:supfac}

As is standard procedure in the \master\ technique~\cite{hivon02},
we noise-debias the spectra by subtracting the mean of
the noise realizations (see Sec.~\ref{sec:noisesims}).
The noise in our maps is so low that this is a relevant
correction only for $BB$, although we do it for all spectra.
(The $BB$ noise debias is $0.006$~$\mu$K$^2$ in the $\ell \approx 75$ band power.)

To determine the response of each observed band power to 
each multipole on the sky we run special simulations
with $\delta$ function spectra input to \synfast\ multiplied
by the average measured beam function.
Taking the mean over many realizations
(to enable the 600 multipoles $\times$ 100 realizations 
per multipole required we do ``short-cut'' 
simulations using the observing matrix $\mathbf{R}$ mentioned 
in Sec.~\ref{sec:matpure} above rather than the usual 
explicit time stream simulations; 
these are empirically found to be equivalent to
high accuracy)
we determine the ``band power window functions'' (BPWF)~\cite{knox99}.
The integral of these functions is the factor by
which each band power has been suppressed by the instrument
beam and all filterings (including the matrix purification).
We therefore divide by these factors and renormalize
the BPWF to unit sum.
This is a variant on the standard \master\ technique.
(We choose to plot the band power values at the weighted mean
of the corresponding BPWF instead of at the nominal
band center.)

One point worth emphasizing is that when comparing the
real data to our simulations
(or jackknife differences thereof) the
noise subtraction and filter or beam corrections have no effect since they
are applied equally to the real data and simulations.
The BPWFs are required to compare the final
band powers to an arbitrary external theoretical model
and are provided with the data release.

The same average measured beam function is used in the signal simulations
and in the BPWF calculation.
In as much as this function does not reflect reality the
real band powers will be under- or overcorrected at
high $\ell$.
We estimate the beam function uncertainty to be equivalent
to a 1.1\% width error on a $31$~arcmin FWHM Gaussian.

\section{Results}
\label{sec:powspecres}

\subsection{Power spectra}
\label{sec:spectra}

\begin{figure*}
\resizebox{\textwidth}{!}{\includegraphics{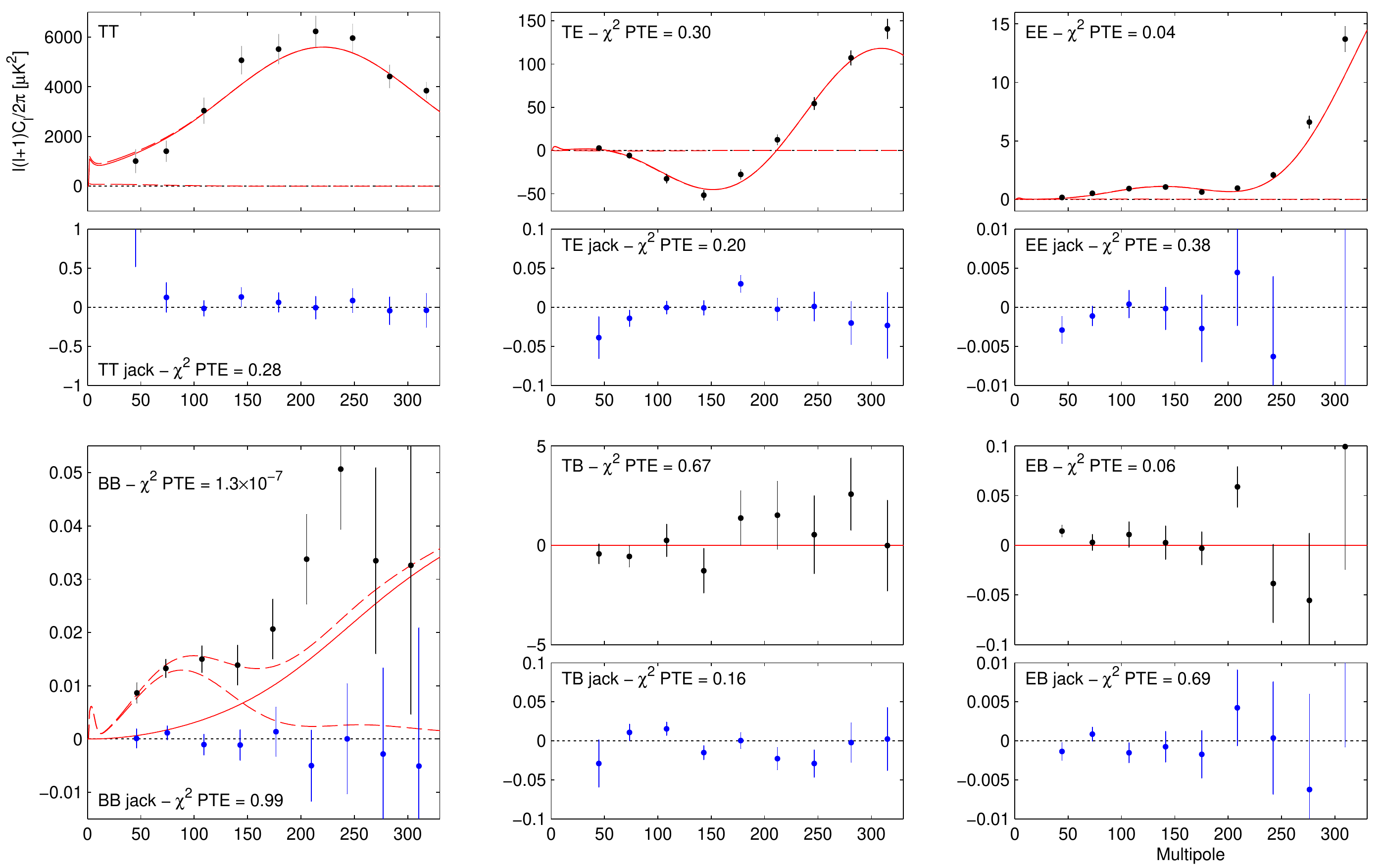}}
\caption{\biceptwo\ power spectrum results for signal (black points)
and temporal-split jackknife (blue points).
The solid red curves show the lensed-\lcdm\ theory expectations while
the dashed red curves show $r=0.2$ tensor spectra and the
sum of both.
The error bars are the standard deviations of the
lensed-\lcdm+noise simulations and hence contain no sample variance
on tensors.
The probability to exceed (PTE) the observed value of a simple
$\chi^2$ statistic is given (as evaluated against the simulations).
Note the very different $y$-axis scales for the jackknife spectra
(other than $BB$).
See the text for additional discussion of the $BB$ spectrum.
(Note that the calibration procedure uses $EB$ to set the overall polarization angle
so $TB$ and $EB$ as plotted above cannot be used to measure astrophysical
polarization rotation---see Sec.~\ref{sec:polrot}.)
}
\label{fig:powspecres}
\end{figure*}

Following the convention
of C10 and B14 we report nine band powers,
each $\approx35$ multipoles wide and spanning
the range $20 < \ell < 340$.
Figure~\ref{fig:powspecres} shows the \biceptwo\ power spectra~\footnote{
These band powers along with all the ancillary data, noise information,
likelihood code, etc. required to
use them are available for download at \url{http://bicepkeck.org/}.}.
With the exception of $BB$ all spectra are consistent with their lensed-\lcdm\ expectation values---the 
probability to exceed (PTE) the observed value of a simple
$\chi^2$ statistic is given on the plot (as evaluated against 
simulations---see Sec.~\ref{sec:jacks}).

$BB$ appears consistent with the lensing expectation at higher
$\ell$, but at lower multipoles there is an excess which
is detected with high signal to noise.
The $\chi^2$ of the data is much too high to allow us to
directly evaluate the PTE of the observed value
under lensed \lcdm\ using the simulations.
We therefore ``amplify'' the Monte Carlo statistics
by resampling band-power values from distributions
fit to the simulated ones.
For the full set of nine band powers shown in the figure
we obtain a PTE of $1.3\times10^{-7}$
equivalent to a significance of $5.3 \sigma$.
Restricting to the first five band powers ($\ell \lesssim 200$)
this changes to $5.2 \sigma$.
We caution against over interpretation of the two high
band powers at $\ell\approx 220$---their joint significance
is $<3\sigma$ (also see Fig.~\ref{fig:speccomp}).

Figure~\ref{fig:powspecres} also shows the temporal-split jackknife---the 
spectrum produced when differencing maps
made from the first and second halves of the data.
The $BB$ excess is not seen in the jackknife, which disfavors
misestimation of the noise debias as the cause
(the noise debias being equally large in jackknife spectra).

\subsection{$E$ and $B$ maps}

\begin{figure*}
\resizebox{\textwidth}{!}{\includegraphics{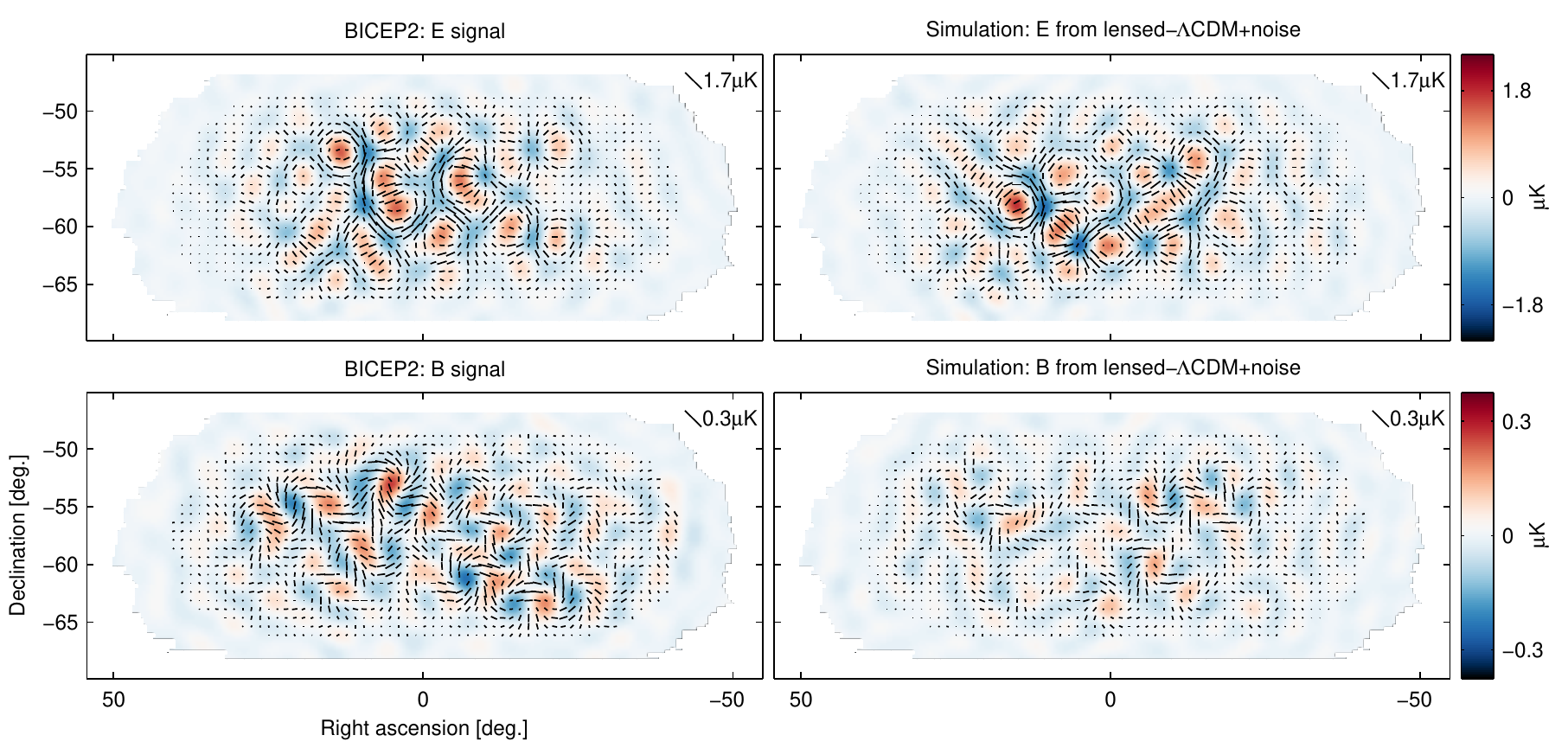}}
\caption{Left: \biceptwo\ apodized \emode\ and \bmode\ maps filtered to $50<\ell<120$.
Right: The equivalent maps for the first of the lensed-\lcdm+noise simulations.
The color scale displays the \emode\ scalar and \bmode\ pseudoscalar patterns
while the lines display the equivalent magnitude and orientation of linear polarization.
Note that excess $B$ mode is detected over lensing+noise with high signal-to-noise
ratio in the map ($s/n>2$ per map mode at $\ell\approx70$).
(Also note that the \emode\ and \bmode\ maps use different
color and length scales.)
}
\label{fig:eb_maps}
\end{figure*}

Once we have the sets of $E$ and $B$ Fourier modes, instead
of collapsing within annuli to form power spectra,
we can instead reinvert to make apodized $E$ and $B$ maps.
In Fig.~\ref{fig:eb_maps} we show such maps
prepared using exactly the same Fourier modes
as were used to construct the spectra shown in
Fig.~\ref{fig:powspecres} filtering to the range $50<\ell<120$.
In comparison to the simulated maps we see
(i) \biceptwo\ has detected $B$ modes with high
signal-to-noise ratio \emph{in the map}, and
(ii) this signal appears to be evenly distributed
over the field, as is the expectation for a cosmological signal, but generally
will not be for a Galactic foreground.

\subsection{Internal consistency tests}
\label{sec:jacks}

We evaluate the consistency of the jackknife spectra
with their \lcdm\ expectations by using a simple $\chi^2$ statistic,
\begin{equation}
\chi^2 = \left( \mathbf{d} - \langle\mathbf{m}\rangle\right)^\mathrm{T} \mathbf{D}^{-1}
\left(\mathbf{d} - \langle\mathbf{m}\rangle\right)
\label{eqn:chisq}
\end{equation}
where $\mathbf{d}$ is the vector of observed band-power values,
$\langle\mathbf{m}\rangle$ is
the mean of the lensed-\lcdm+noise simulations (except where
alternative signal models are considered), and $\mathbf{D}$
is the band-power covariance matrix as evaluated from those 
simulations.
(Because of differences in sky coverage between the two halves of
a jackknife split, in conjunction with filtering, the expectation
value of some of the jackknifes is not quite zero---hence we always
evaluate $\chi^2$ versus the mean of the simulations.
Because the BPWF overlap slightly adjacent band powers are $\lesssim10$\%
correlated.
We zero all but the main and first off-diagonal elements of $\mathbf{D}$
as the other elements are not measured above noise given the
limited simulation statistics.)
We also compute $\chi^2$ for each of the simulations
(recomputing $\mathbf{D}$ each time, excluding that simulation)
and take the probability to exceed (PTE) the observed value
versus the simulated distribution.
In addition to $\chi^2$ we compute the sum of normalized
deviations,
\begin{equation}
\chi = \sum_i \frac{d_i - \langle m_i \rangle}{\sigma_{m_i}}
\end{equation}
where the $d_i$ are the observed band-power values and $\langle m_i \rangle$
and $\sigma_{m_i}$ are the mean and standard deviation of the
lensed-\lcdm+noise simulations.
This statistic tests for sets of band powers which are
consistently all above or below the expectation.
Again we evaluate the PTE of the observed value against the
distribution of the simulations.

\begin{figure}
\resizebox{\columnwidth}{!}{\includegraphics{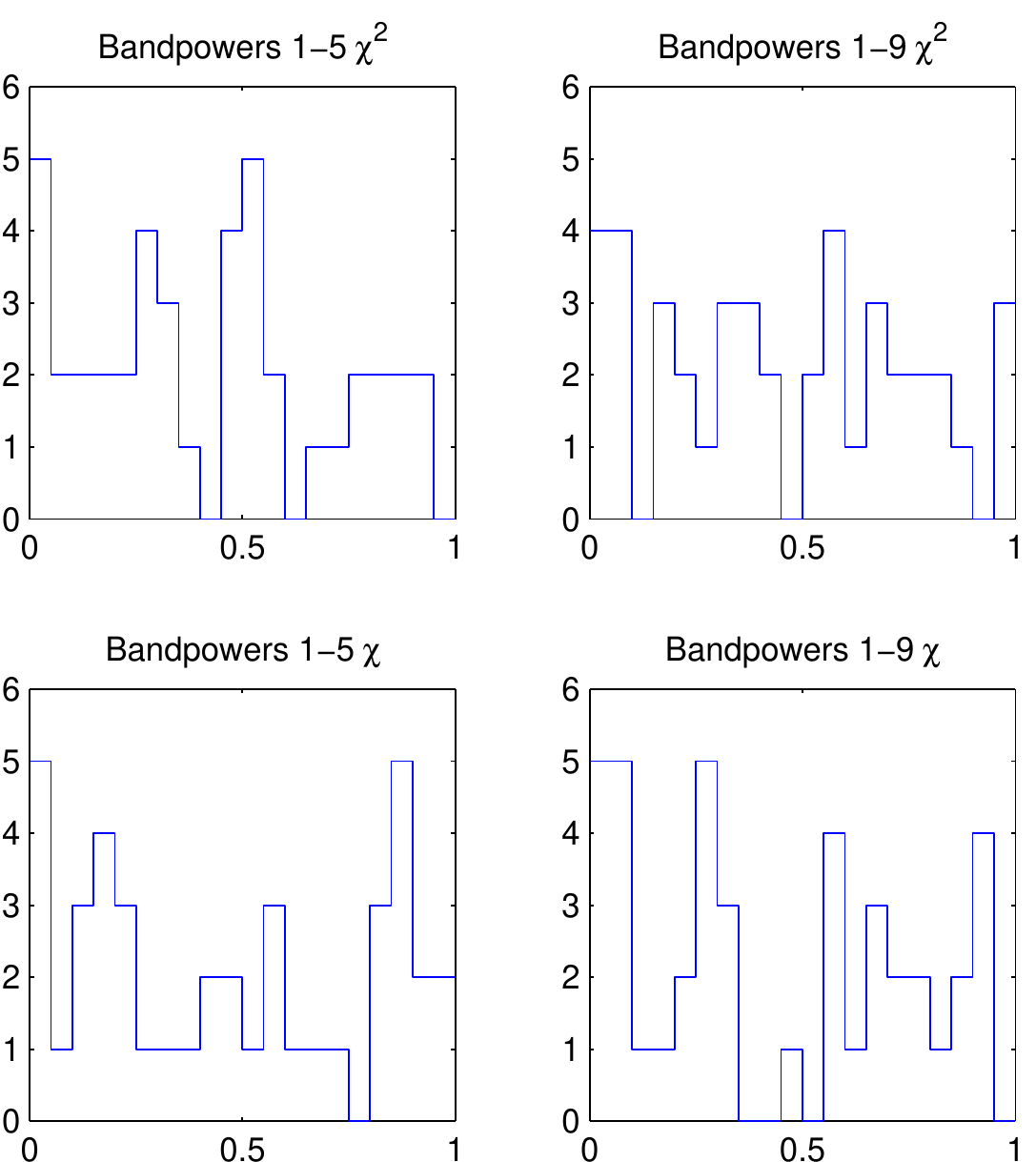}}
\caption{Distributions of the jackknife $\chi^2$ and $\chi$ PTE
values over the 14 tests and three spectra given in Table~\ref{tab:ptes}.
These distributions are consistent with uniform.
}
\label{fig:ptedist}
\end{figure}

We evaluate these statistics both for the full set of nine band powers
(as in C10 and B14), and also for the lower five
of these corresponding to the multipole range of greatest
interest ($20<\ell<200$).
Numerical values are given in Table~\ref{tab:ptes}
and the distributions are plotted in Fig.~\ref{fig:ptedist}.
Since we have 500 simulations the minimum observable nonzero 
value is 0.002.
Most of the $TT$, $TE$, and $TB$ jackknifes pass,
but following C10 and B14 we omit them from formal consideration
(and they are not included in the table and figure).
The signal-to-noise ratio in $TT$ is $\sim10^4$ so tiny differences in absolute
calibration between the data subsets can cause jackknife failure,
and the same is true to a lesser extent for $TE$ and $TB$.
Even in $EE$ the signal-to-noise is approaching $\sim10^3$
(500 in the $\ell\approx 110$ bin)
and in fact most of the low values in the table are in $EE$.
However, with a maximum signal-to-noise ratio of $\lesssim 10$ in $BB$
such calibration differences are not a concern.
All the $BB$ (and $EB$) jackknifes are seen to pass,
with the 112 numbers in Table~\ref{tab:ptes}
having one greater than 0.99, one less than 0.01 and a
distribution consistent with uniform.
Note that the four test statistics for each spectrum and
jackknife are correlated this must be taken into account
when assessing uniformity.

\begingroup
\squeezetable
\begin{table}
\caption{Jackknife PTE values from $\chi^2$ and $\chi$ (sum of deviation) tests \label{tab:ptes}}
\begin{ruledtabular}
\begin{tabular}{l c c c c}
Jackknife & Bandpowers & Bandpowers & Bandpowers & Bandpowers \\
& 1--5 $\chi^2$ & 1--9 $\chi^2$ & 1--5 $\chi$ & 1--9 $\chi$ \\
\hline
\multicolumn{5}{l}{Deck jackknife} \\
$EE$ & 0.046 & 0.030 & 0.164 & 0.299 \\ 
$BB$ & 0.774 & 0.329 & 0.240 & 0.082 \\ 
$EB$ & 0.337 & 0.643 & 0.204 & 0.267 \\ 
\multicolumn{5}{l}{Scan dir jackknife} \\
$EE$ & 0.483 & 0.762 & 0.978 & 0.938 \\ 
$BB$ & 0.531 & 0.573 & 0.896 & 0.551 \\ 
$EB$ & 0.898 & 0.806 & 0.725 & 0.890 \\ 
\multicolumn{5}{l}{Temporal split jackknife} \\
$EE$ & 0.541 & 0.377 & 0.916 & 0.938 \\ 
$BB$ & 0.902 & 0.992 & 0.449 & 0.585 \\ 
$EB$ & 0.477 & 0.689 & 0.856 & 0.615 \\ 
\multicolumn{5}{l}{Tile jackknife} \\
$EE$ & 0.004 & 0.010 & 0.000 & 0.002 \\ 
$BB$ & 0.794 & 0.752 & 0.565 & 0.331 \\ 
$EB$ & 0.172 & 0.419 & 0.962 & 0.790 \\ 
\multicolumn{5}{l}{Azimuth jackknife} \\
$EE$ & 0.673 & 0.409 & 0.126 & 0.339 \\ 
$BB$ & 0.591 & 0.739 & 0.842 & 0.944 \\ 
$EB$ & 0.529 & 0.577 & 0.840 & 0.659 \\ 
\multicolumn{5}{l}{Mux col jackknife} \\
$EE$ & 0.812 & 0.587 & 0.196 & 0.204 \\ 
$BB$ & 0.826 & 0.972 & 0.293 & 0.283 \\ 
$EB$ & 0.866 & 0.968 & 0.876 & 0.697 \\ 
\multicolumn{5}{l}{Alt deck jackknife} \\
$EE$ & 0.004 & 0.004 & 0.070 & 0.236 \\ 
$BB$ & 0.397 & 0.176 & 0.381 & 0.086 \\ 
$EB$ & 0.150 & 0.060 & 0.170 & 0.291 \\ 
\multicolumn{5}{l}{Mux row jackknife} \\
$EE$ & 0.052 & 0.178 & 0.653 & 0.739 \\ 
$BB$ & 0.345 & 0.361 & 0.032 & 0.008 \\ 
$EB$ & 0.529 & 0.226 & 0.024 & 0.048 \\ 
\multicolumn{5}{l}{Tile and deck jackknife} \\
$EE$ & 0.048 & 0.088 & 0.144 & 0.132 \\ 
$BB$ & 0.908 & 0.840 & 0.629 & 0.269 \\ 
$EB$ & 0.050 & 0.154 & 0.591 & 0.591 \\ 
\multicolumn{5}{l}{Focal plane inner or outer jackknife} \\
$EE$ & 0.230 & 0.597 & 0.022 & 0.090 \\ 
$BB$ & 0.216 & 0.531 & 0.046 & 0.092 \\ 
$EB$ & 0.036 & 0.042 & 0.850 & 0.838 \\ 
\multicolumn{5}{l}{Tile top or bottom jackknife} \\
$EE$ & 0.289 & 0.347 & 0.459 & 0.599 \\ 
$BB$ & 0.293 & 0.236 & 0.154 & 0.028 \\ 
$EB$ & 0.545 & 0.683 & 0.902 & 0.932 \\ 
\multicolumn{5}{l}{Tile inner or outer jackknife} \\
$EE$ & 0.727 & 0.533 & 0.128 & 0.485 \\ 
$BB$ & 0.255 & 0.086 & 0.421 & 0.036 \\ 
$EB$ & 0.465 & 0.737 & 0.208 & 0.168 \\ 
\multicolumn{5}{l}{Moon jackknife} \\
$EE$ & 0.499 & 0.689 & 0.481 & 0.679 \\ 
$BB$ & 0.144 & 0.287 & 0.898 & 0.858 \\ 
$EB$ & 0.289 & 0.359 & 0.531 & 0.307 \\ 
\multicolumn{5}{l}{Differential pointing best or worst} \\
$EE$ & 0.317 & 0.311 & 0.868 & 0.709 \\ 
$BB$ & 0.114 & 0.064 & 0.307 & 0.094 \\ 
$EB$ & 0.589 & 0.872 & 0.599 & 0.790 \\ 
\end{tabular}
\end{ruledtabular}
\end{table}
\endgroup

To form the jackknife spectra we difference the maps made
from the two halves of the data split, divide by two, and
take the power spectrum.
This holds the power spectrum amplitude of a contribution which
is uncorrelated in the two halves (such as noise) constant, while a fully
correlated component (such as sky signal) cancels perfectly.
The amplitude of a component which appears only in one half will stay
the same under this operation as it is in the fully coadded map 
and the apparent signal-to-noise will also stay the same.
For a noise-dominated experiment
this means that jackknife tests can only limit potential contamination
to a level comparable to the noise uncertainty.
However, the $BB$ bandpowers shown in Fig.~\ref{fig:powspecres}
have signal-to-noise as high as 10.
This means that jackknife tests are extremely powerful in our case---the 
reductions in power which occur in the jackknife spectra
are empirical proof that the \bmode\ pattern on the sky
is highly correlated between all data splits considered.

We have therefore conducted an unusually large number of
jackknife tests trying to imagine data splits which
might conceivably contain differing contamination.
Here we briefly describe each of these:

\biceptwo\ observed at deck angles of $68\deg$, $113\deg$, $248\deg$ and $293\deg$.
We can split these in two ways without losing the ability
to make $Q$ and $U$ maps (see Sec.~\ref{sec:acctomaps}).
The {\it deck} jackknife is defined as $68\deg$ and $113\deg$ vs $248\deg$ and $293\deg$
while the {\it alt.\ deck} jackknife is $68\deg$ and $293\deg$ vs $113\deg$ and $248\deg$.
Uniform differential pointing averages down in a
coaddition of data including an equal mix of $180\deg$ complement
angles, but it will be amplified in either of these jackknifes
(as we see in our simulations). The fact that we are passing 
these jackknifes indicates that residual beam systematics 
of this type are subdominant after deprojection. 

The {\it temporal-split}
simply divides the data into two equal weight parts sequentially.
Similarly, but at the opposite end of the time scale
range, we have the {\it scan direction} jackknife,
which differences maps made from the right and left going
half scans, and is sensitive to errors in the detector
transfer function.

The {\it azimuth} jackknife differences data taken over
different ranges of telescope azimuth angle---i.e.,
with different potential contamination
from fixed structures or emitters on the ground.
A related category is the {\it moon} jackknife, which
differences data taken when the moon is above and below
the horizon.

A series of jackknifes tests if the signal originates
in some subset of the detector pairs.
The {\it tile} jackknife tests tiles 1 and 3 vs 2 and 4 (this combination
being necessary to get reasonable coverage in the $Q$ and $U$ maps).
Similarly the {\it tile inner or outer} and {\it tile top or bottom} jackknifes
are straightforward.
The {\it focal plane inner or outer} does as stated for the entire
focal plane and is a potentially powerful test for imperfections
which increase radially.
The {\it mux row} and {\it mux column} jackknifes test for
systematics originating in the readout system.

The {\it tile and deck} jackknife tests for a possible effect
coming from always observing a given area of sky with
detectors the ``same way up,'' although due to the small
range of the elevation steps it is limited to a small
sky area.

Finally we have performed one test based on beam nonideality
as observed in external beam map runs.
The {\it differential pointing best or worst} jackknife differences the best and
worst halves of the detector pairs as selected by that
metric.

See the Systematics Paper for a full description of the
jackknife studies.

\section{Systematic Uncertainties}
\label{sec:sysuncer}

Within the simulation-calibrated analysis
framework described above we are free to perform any arbitrary
filtering of the data which may be necessary to render the
results insensitive to particular systematics.
However, as such mode removal increases the
uncertainty of the final band powers, we clearly
wish to filter only systematics which might induce false
$B$ mode at relevant levels.
Moreover, it may not be computationally feasible to construct
simple time stream templates for some potential systematics.
Therefore once we have made our selection as to which filterings to perform
we must then estimate the residual contamination
and either subtract it or show it to be negligible.

To guide our selection of mode removal we have two main
considerations.
First, we can examine jackknifes of the type described in Sec.~\ref{sec:jacks}
above---reduction in failures with increasing mode removal
may imply that a real systematic effect is present.
Second, and as we will see below more powerfully, we can
examine external calibration data (principally beam maps)
to directly calculate the false $B$ mode expected
from specific effects.

\subsection{Simulations using observed per-channel beam shapes}
\label{sec:beammapsim}

As described in the Instrument Paper we have made extremely
high signal-to-noise \textit{in situ} measurements of the far-field
beam shape of each channel.
Fitting these beams to elliptical Gaussians we obtain
differential parameters that correlate well with the mean value
of the deprojection coefficients from Sec.~\ref{sec:deproj}.
One may then ask whether it would be better to subtract
rather than deproject.
In general it is more conservative to deproject
as this (i) allows for the possibility that the coefficients
are changing with time, and (ii) is guaranteed to completely
eliminate the effect in the mean, rather than leaving a residual
bias due to noise on the subtraction coefficients.

\begin{figure*}
\begin{center}
\resizebox{\columnwidth}{!}{\includegraphics{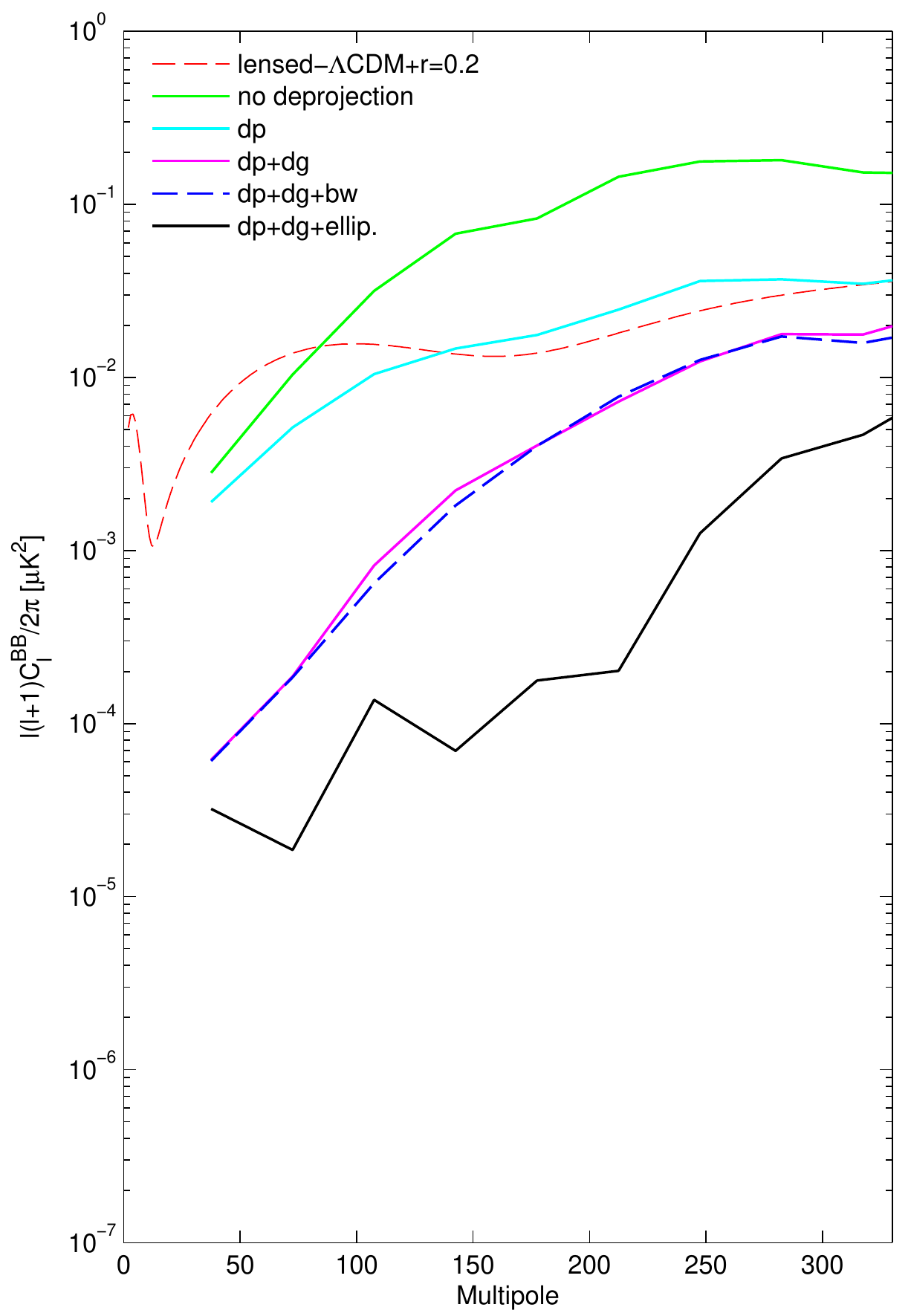}}
\resizebox{\columnwidth}{!}{\includegraphics{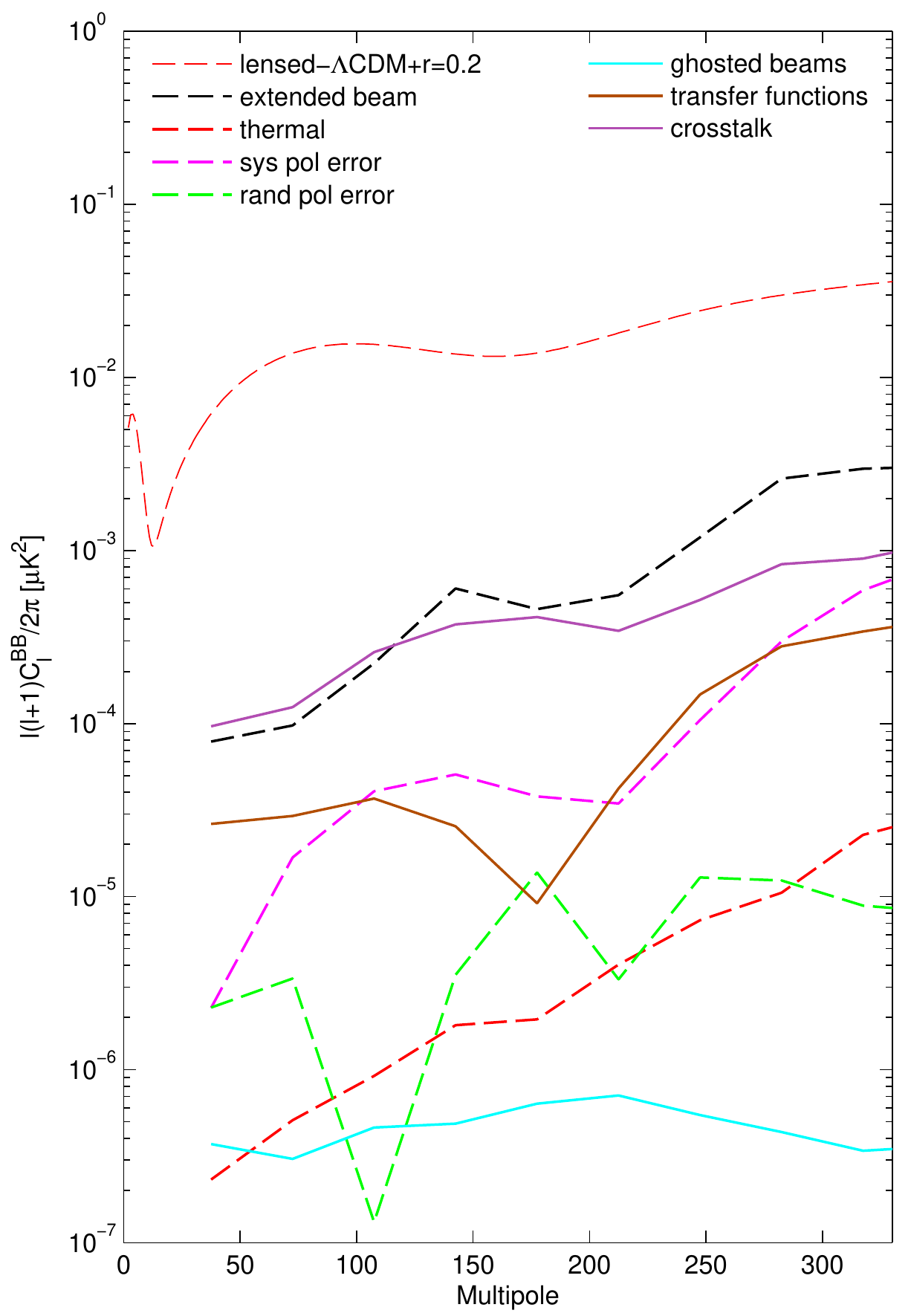}}
\end{center}
\caption{Left: $BB$ spectra from $T$-only input simulations using
the measured per channel beam shapes
compared to the lensed-\lcdm+$r=0.2$ spectrum.
From top to bottom the curves are (i) no deprojection,
(ii) deprojection of differential pointing only (dp),
(iii) deprojection of differential pointing and differential gain of the
detector pairs (dp+dg),
(iv) adding deprojection of differential beam width (dp+dg+bw), and
(v) differential pointing, differential gain, and differential
ellipticity (dp+dg+ellip).
Right: Estimated levels of other systematics
as compared to the lensed-\lcdm+$r=0.2$ spectrum.
Solid lines indicate expected contamination.
Dashed lines indicate upper limits. All systematics are comparable to or smaller 
than the extended beam mismatch upper limit.
}
\label{fig:sys.}
\end{figure*}

We use the per-channel beam maps as inputs to special $T$-only input simulations
and measure the level of $T$ to $B$ mixing while varying
the set of beam modes being deprojected.
The beam maps do not provide a good estimate of differential gain
so we substitute estimates which come
from a per-channel variant of the absolute calibration
procedure mentioned in Sec.~\ref{sec:acctomaps} above.
The left panel of Fig.~\ref{fig:sys.} shows \bmode\ power spectra
from these simulations under the following
deprojections (i) none, (ii) differential pointing only,
(iii) differential pointing and differential gain,
(iv) differential pointing, differential gain, and differential beam width,
and (v) differential pointing, differential gain, and differential
ellipticity.

We see that differential pointing has the largest
effect and so to be conservative we choose
to deproject it.
Differential gain is also seen to be a significant
effect and we again deproject it---we lack
independent subtraction coefficients, and it might
plausibly be time variable.
Differential beam width is a negligible effect and we
do not deproject it.
Differential ellipticity is also a small effect.
We find in the simulations that deprojection of differential ellipticity
interacts with real $TE$ correlation in a complex manner
slightly distorting the $TE$ spectrum.
We therefore choose to subtract this effect by fixing the
coefficients to their beam map derived values
in Sec.~\ref{sec:deproj}.
Whether differential ellipticity is deprojected or
subtracted makes no significant difference to any
of the spectra other than $TE$.
Finally, we make a small correction for the
undeprojected residual by subtracting the final curve
in the left panel of Fig.~\ref{fig:sys.} from the
results presented in Sec.~\ref{sec:powspecres}.
(The correction is equivalent to $r=0.001$.)
We also increase the band power fluctuation to reflect 
the postcorrection upper limit on extended beam mismatch
shown in the right panel of Fig.~\ref{fig:sys.}.
See the Systematics Paper for details.

\subsection{Overall polarization rotation}
\label{sec:polrot}

Once differential ellipticity has been corrected we
notice that an excess of $TB$ and $EB$ power
remains at $\ell>200$ versus the \lcdm\ expectation.
The spectral form of this power is consistent
with an overall rotation of the polarization angle
of the experiment.
While the detector-to-detector relative angles have been measured
to differ from the design values by $<0.2\deg$ we currently
do not have an accurate external measurement of the
overall polarization angle.
We therefore apply a rotation of $\sim1\deg$ to the final $Q$ and $U$ maps
to minimize the $TB$ and $EB$ power~\cite{KSY,Kaufman13}.
We emphasize that this has a negligible
effect on the $BB$ bandpowers at $\ell<200$.
(The effect is $1.5\times10^{-3}$~$\mu$K$^2$ at $\ell \sim 130$
and decreasing to lower $\ell$.)

\subsection{Other possible systematics}

Many other systematics can be proposed as possibly leading to false $B$ modes at
a relevant level.
Some possible effects will produce jackknife failure before
contributing to the nonjackknife \bmode\ power at a relevant level.
Limits on
others must be set by external data or other considerations.
Any azimuth fixed effect, such as magnetic
pickup, is removed by the scan-synchronous template
removal mentioned in Secs.~\ref{sec:obsobs} and~\ref{sec:mapmaking}.

We have attempted an exhaustive consideration of all possible effects---a
brief summary will be given here with the details deferred to
the Systematics Paper.
The right panel of Fig.~\ref{fig:sys.} shows estimated levels
of, or upper limits on, contamination from extended beam mismatch
after the undeprojected residual correction,
thermal drift in the focal plane, systematic
polarization angle miscalibration, randomized polarization angle miscalibration,
ghost beams, detector transfer function mismatch, and crosstalk.
The upper limit for extended beam mismatch is the $1\sigma$ uncertainty
on contamination predicted from beam map simulations identical to those
described in Sec.~\ref{sec:beammapsim} but using a larger region of the beam.
(Note that this will include beam or
beamlike effects which are present in the
beam mapping runs, including crosstalk and side lobes at $\lesssim 4\deg$.)
For systematic polarization angle miscalibration it is the
level at which such an error would produce a detectable $TB$ signal with $95\%$
confidence.
For randomized polarization angle miscalibration, it is the leakage
we would incur from assuming nominal polarization angles, i.e., no ability to
measure per-pair relative polarization angles. For thermal drift, it is the
noise floor set by the sensitivity of the thermistors that monitor focal
plane temperature.

\section{Foreground Projections}
\label{sec:foregrounds}

Having provided evidence that the detected \bmode\ signal is not an instrumental
artifact, we now consider whether it might be due to a
Galactic or extragalactic foreground.
At low or high frequencies Galactic synchrotron and polarized-dust 
emission, respectively, are the dominant foregrounds.
The intensity of both falls rapidly with increasing Galactic latitude
but dust emission falls faster.
The equal amplitude crossover frequency therefore rises
to $\gtrsim 100$~GHz in the cleanest regions~(\cite{dunkley08}, Fig. 10).
The \biceptwo\ field is centered on Galactic coordinates
$(l,b)=(316\deg, -59\deg)$ and was originally selected on the basis of
exceptionally low contrast in the FDS dust maps~\cite{finkbeiner99}.
In these unpolarized maps such ultraclean regions are very special---at 
least an order of magnitude cleaner than the
average $b>50\deg$ level.

Foreground modeling involves extrapolating maps
taken at lower or higher frequencies to the CMB observation
band, and there are inevitably uncertainties.
Many previous studies have been conducted
and projections made---see, for instance, Dunkley \textit{et al.}~\cite{dunkley08}, and references therein.
Such previous studies have generically predicted 
levels of foreground \bmode\ contamination
in clean high latitude regions equivalent to $r\lesssim 0.01$---well 
below that which we observe---although 
they note considerable uncertainties.

\subsection{Polarized dust projections}

The main uncertainty in foreground modeling is currently the lack
of a polarized dust map. (This will be alleviated soon by
the next \planck\ data release.)
In the meantime we have therefore investigated a number of existing
models using typical or default assumptions for polarized
dust, and have formulated a new one.
A brief description of each model is as follows:

\textit{FDS}: Model 8~\cite{finkbeiner99},
scaled with a uniform polarization fraction of 5\%, is a commonly
used all-sky baseline model (e.g., \cite{page07,dunkley08}).  We set $Q=U$.

\textit{BSS}: Bisymmetric spiral (BSS) model of the Galactic magnetic
field~\cite{odea11,contaldiweb}.
The polarization fraction varies across the sky in this model; 
by default it is scaled to match the 3.6\% all-sky average reported by \wmap~\cite{kogut07},
giving a mean and standard deviation in the \biceptwo\ field of
$(5.7\pm0.7)$\%.

\textit{LSA}: Logarithmic spiral arm (LSA) model of the Galactic magnetic
field~\cite{odea11,contaldiweb}.
The polarization fraction varies across the sky in this model; 
by default it is also scaled to match the 3.6\% all-sky average reported by \wmap~\cite{kogut07},
giving a mean and standard deviation in the \biceptwo\ field of
$(5.0\pm0.3)$\%.

\textit{PSM}: Planck sky model (PSM)~\cite{delabrouille13} version
1.7.8, run as a ``Prediction'' with default settings, including
15\% dust intrinsic polarization fraction~\footnote{
\url{http://www.apc.univ-paris7.fr/~delabrou/PSM/psm.html}}.
In this model, the intrinsic polarization fraction is reduced by averaging
over variations along each line of sight.
The resulting polarization fraction varies across the sky; 
its mean and standard deviation in the \biceptwo\ field are
$(5.6\pm0.8)$\%.

\textit{DDM1}: ``Data driven model 1'' (DDM1) constructed from publicly available
\planck\ data products.
The \planck\ dust model map at 353~GHz is
scaled to 150~GHz assuming a constant emissivity value
of 1.6 and a constant temperature of 19.6~K~\cite{planckXI}.
A nominal uniform 5\% sky polarization fraction is assumed, and
the polarization angles are taken from the PSM.
This model will be biased down due to the lack of spatial
fluctuation in the polarization fraction and angles, but
biased up due to the presence of instrument noise and (unpolarized)
cosmic infrared background anisotropy in the \planck\ dust 
model~\footnote{
In the preprint version of this paper an additional DDM2 model
was included based on information taken from \planck\ conference talks.
We noted the large uncertainties on this and the other dust models presented.
In the \planck\ dust polarization
paper~\cite{planckiXIX} which has since appeared
the maps have been masked to include only regions ``where the systematic
uncertainties are small, and where the dust signal dominates total emission.''
This mask excludes our field.
We have concluded the information used for the DDM2 model
has unquantifiable uncertainty.
We look forward to performing a cross-correlation analysis
against the \planck\ 353~GHz polarized maps in a future publication.
}.

\begin{figure}
\resizebox{\columnwidth}{!}{\includegraphics{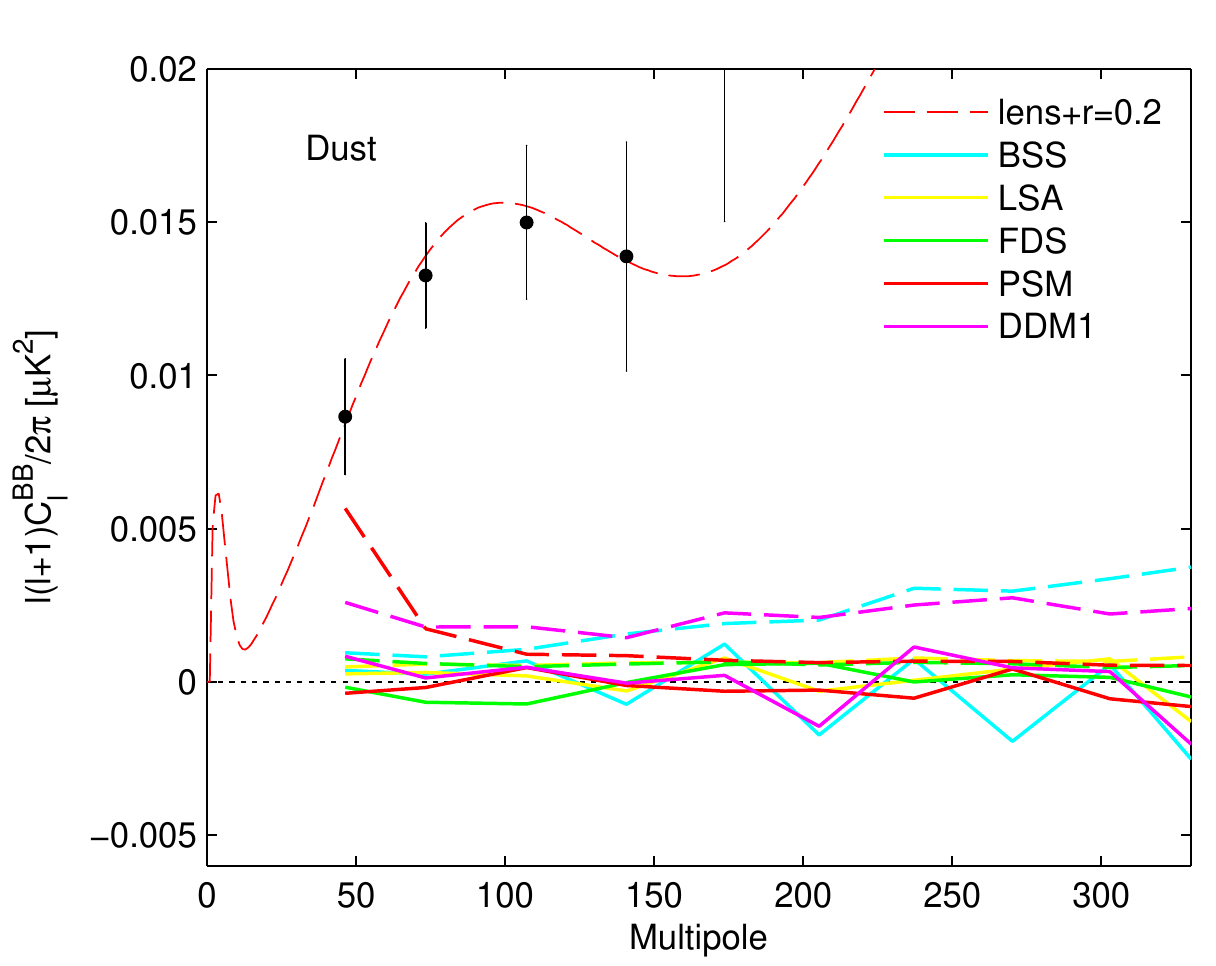}}
\resizebox{\columnwidth}{!}{\includegraphics{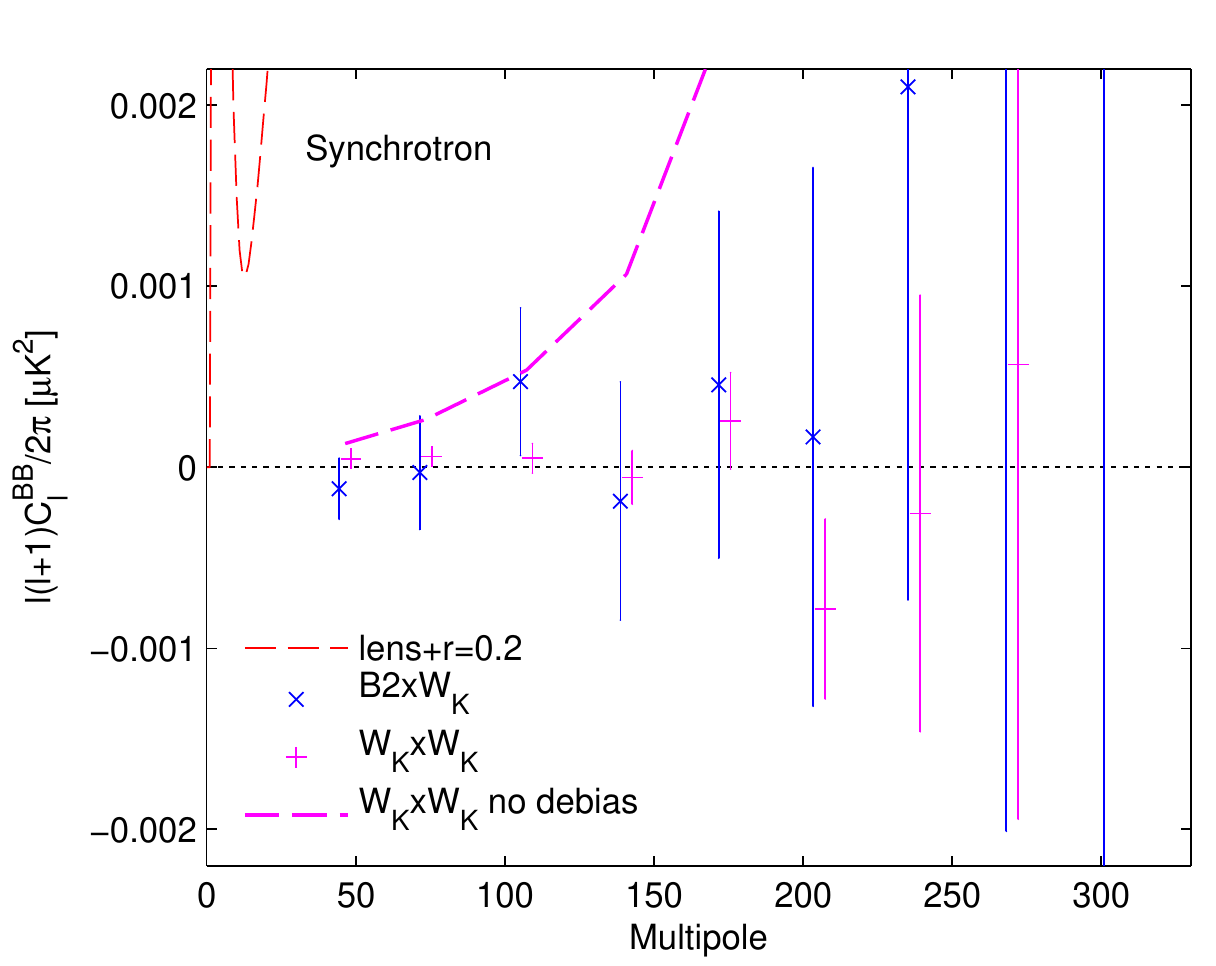}}
\caption{Upper: Polarized dust foreground projections for our field using
various models available in the literature, and a new
one formulated using the information officially available from \planck.
Dashed lines show autospectra of the models, while
solid lines show cross spectra between the models and the \biceptwo\
maps.
The \biceptwo\ auto spectrum from Fig.~\ref{fig:powspecres} is
also shown with the lensed-\lcdm+$r=0.2$ spectrum.
Lower: Polarized synchrotron constraints for our field using
the \wmap\ \textit{K} band (23~GHz) maps projected to 150~GHz using the mean
spectral index within our field ($\beta=-3.3$) from \wmap.
The blue points with error bars show the cross spectrum between
the \biceptwo\ and \wmap\ maps, with the uncertainty estimated
from cross spectra against simulations of the \wmap\ noise.
The magenta points with error bars and the dashed curve 
show the \wmap\ auto spectrum with and without noise debias.
See the text for further details.
}
\label{fig:foregroundproj}
\end{figure}

All of the models except FDS make explicit predictions of the actual
polarized dust pattern in our field.
We can therefore search for a correlation between the models and
our signal by taking cross spectra against the \biceptwo\ maps.
The upper panel of Fig.~\ref{fig:foregroundproj} shows the resulting $BB$
auto and cross spectra---the 
autospectra are all below the level of
our observed signal and no significant cross correlation is found.
[The cross spectra between each model and real data are consistent
with the cross spectra between that model and (uncorrelated)
lensed-LCDM+noise simulations.]
We note that the lack of cross-correlation can be interpreted as
due to limitation of the models.
To produce a power level from DDM1 auto comparable to the
observed excess signal would require one to assume a uniform polarization fraction of
$\sim13$\%.  While this is well above typically assumed values, models
are not yet well-enough constrained by external public data to exclude
the possibility of emission at this level.

\subsection{Synchrotron}
\label{sec:sync}

To constrain the level of Galactic synchrotron in our field we take the
\wmap\ \textit{K}-band (23~GHz) map, extrapolate it to 150~GHz, reobserve with
our simulation pipeline, and take
the cross spectrum against the \biceptwo\ maps,
with appropriate \biceptwo\ filtering and \wmap\ beam correction.
In our field and at angular scales of $\ell>30$ the \wmap\ \textit{K}-band
maps are noise dominated.
We therefore also make noise realizations and take cross spectra with
these to assess the uncertainty.
The lower panel of Fig.~\ref{fig:foregroundproj} shows the resulting
cross spectrum and its uncertainty.
Using the MCMC Model f spectral index map provided by \wmap~\cite{bennett13} we obtain
a mean value within our field of $\beta=-3.3 \pm 0.16$.
For this value, the resulting cross spectrum implies a contribution to our
$r$ constraint (calculated as in Sec.~\ref{sec:parcons}) 
equivalent to $r_{sync,150}=0.0008 \pm 0.0041$,
while for a more conservative $\beta=-3.0$, $r_{sync,150}=0.0014 \pm 0.0071$.
In contrast to analysis with the models of polarized dust, 
cross spectra with the official \wmap\ polarized maps
can be confidently expected to provide an unbiased estimate
of signal correlated with synchrotron for a given spectral index,
with a quantified uncertainty.
Note that the assumed spectral index only enters as the first power
in these \biceptwo$\times$\wmap$_K$ cross spectral constraints, and the uncertainty
depends only weakly on the model for \wmap\ noise.
The \wmap$_K$ auto spectrum,
if de-biased for noise, implies
even tighter constraints on the synchrotron contribution to
our $r$ parameter: for $\beta=-3.3$, $r_{sync,150}=0.001 \pm 0.0006$, 
or for $\beta=-3.0$, $r_{sync,150}=0.003 \pm 0.002$, although
these have a somewhat greater dependence on assumptions about \wmap\ noise
levels and the spectral index.

\subsection{Point sources}

Extragalactic point sources might also potentially
be a concern.
Using the 143~GHz fluxes for the sources in our field
from the \planck\ catalog~\cite{planckXXVIII}, together with polarization
information from ATCA~\cite{massardi11} we find that the contribution
to the $BB$ spectrum is equivalent to $r\approx0.001$.
This is consistent with the projections of Battye \textit{et al.}~\cite{battye10}.

\section{Cross Spectra}
\label{sec:cross_spectra}

\subsection{Cross spectra with \bicepone}
\label{sec:b1xspec150}

\bicepone\ observed essentially the same field as \biceptwo\ from 2006 to 2008.
While a very similar instrument in many ways the focal
plane technology of \bicepone\ was entirely
different, employing horn-fed PSBs read out via neutron transmutation-doped (NTD)
germanium thermistors (see T10 for details). The high-impedance NTD devices and
readouts have different susceptibility to microphonic pickup and magnetic
fields, and the shielding of unwanted RFI and EMI was significantly different from
that of \biceptwo.
The beam systematics were also quite different with a more conservative
edge taper and smaller observed pair centroid offsets (see T10
and the Instrument Paper).
\bicepone\ had detectors at both 100 and 150~GHz.

\begin{figure}
\resizebox{\columnwidth}{!}{\includegraphics{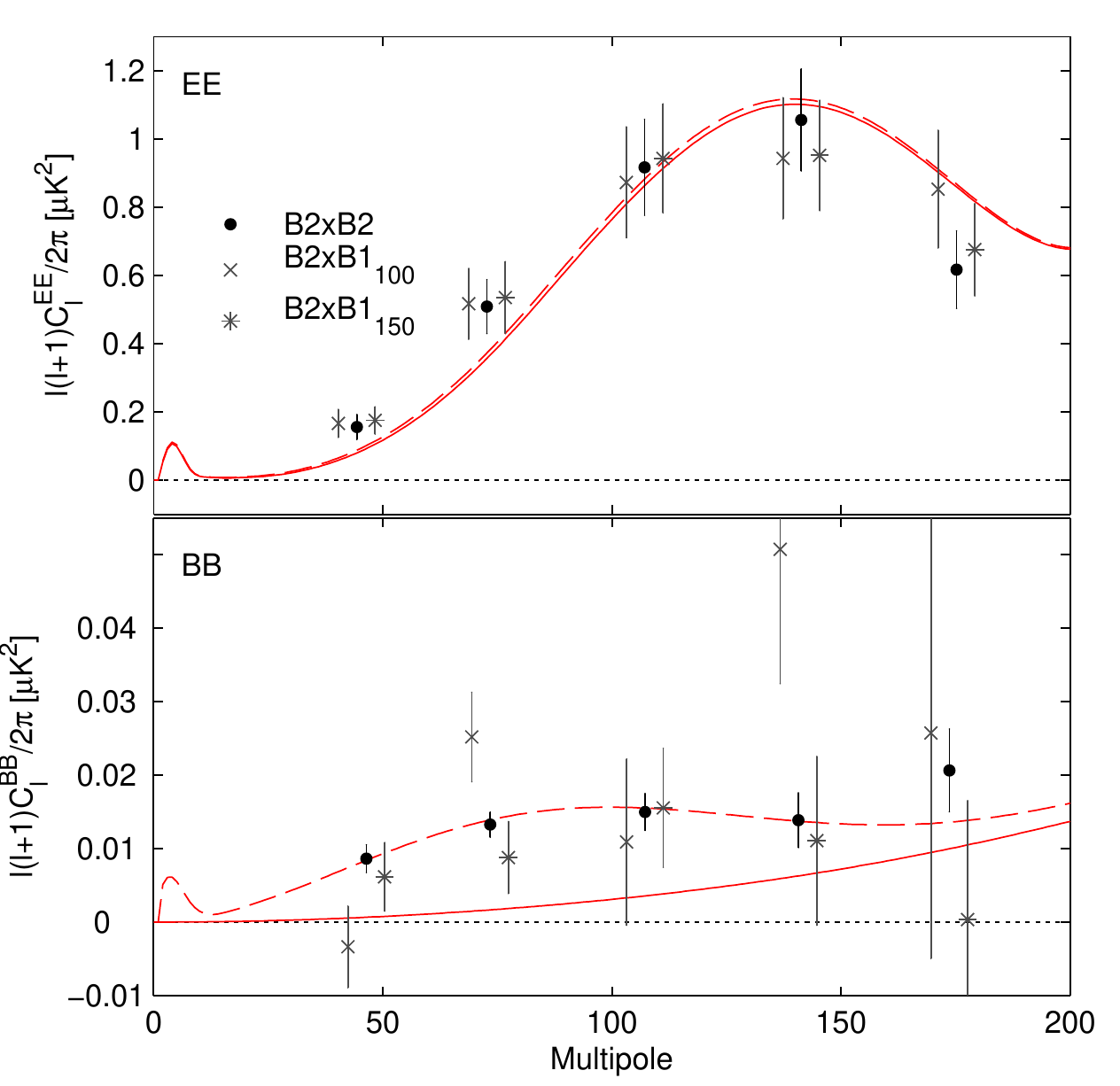}}
\caption{The \biceptwo\ $EE$ and $BB$ auto spectra
(as shown in Fig.~\ref{fig:powspecres}) compared
to cross spectra between \biceptwo\ and the 100
and 150~GHz maps from \bicepone.
The error bars are the standard deviations of the
lensed-\lcdm+noise simulations and hence contain no sample variance
on tensors.
(For clarity the cross spectrum points are offset horizontally.)
}
\label{fig:b2xb1}
\end{figure}

Figure~\ref{fig:b2xb1} compares the \biceptwo\ $EE$ and $BB$ auto
spectra with cross spectra taken against the 100 and
150~GHz maps from \bicepone.
For $EE$ the correlation is extremely strong, which simply
confirms that the mechanics of the process are working as expected.
For $BB$ the signal-to-noise is of course much lower, but there appear to be
positive correlations.
To test the compatibility of the $BB$ auto and cross spectra we take
the differences and compare to the differences
of lensed-\lcdm+noise+$r=0.2$ simulations (which share common input skies).
(For all spectral difference tests we compare against
lensed-\lcdm+noise+$r=0.2$ simulations as the cross terms between signal
and noise increase the variance even for perfectly common sky coverage.)
Using bandpowers 1--5 the $\chi^2$ and $\chi$
PTEs are midrange, indicating
that the spectra are compatible to within the noise.
(This is also true for $EE$.)

To test for evidence of excess power over the base lensed-\lcdm\
expectation we calculate the
$BB$ $\chi^2$ and $\chi$ statistics against this model.
The \biceptwo$\times$\bicepone$_{150}$
spectrum has PTEs of $0.37$ and $0.05$ respectively, while
the \biceptwo$\times$\bicepone$_{100}$ spectrum has
PTEs of 0.005 and 0.001.
The latter corresponds to a $\approx 3\sigma$ detection of excess power.
While it may seem surprising that one cross spectrum shows a stronger
detection than the other, it turns out not to be unusual in
lensed-\lcdm+noise+$r=0.2$ simulations.
(Compared to such lensed-\lcdm+noise+$r=0.2$ simulations,
$\chi^2$ and $\chi$ PTEs are 0.92 and 0.74 for
\biceptwo$\times$\bicepone$_{150}$ and 0.18 and 0.23 for
\biceptwo$\times$\bicepone$_{100}$.
These simulations also indicate that the \biceptwo$\times$\bicepone$_{150}$
and \biceptwo$\times$\bicepone$_{100}$ values are only weakly correlated.
Therefore if $r=0.2$ is the true underlying model then the
observed \biceptwo$\times$\bicepone$_{150}$ $\chi^2$ and $\chi$
values appear to be modest downward fluctuations and the
\biceptwo$\times$\bicepone$_{100}$ values modest upward fluctuations---but 
they are compatible.)

\subsection{Spectral index constraint}
\label{sec:specindcons}

We can use the \biceptwo\ auto and \biceptwo$\times$\bicepone$_{100}$
spectra shown in Fig.~\ref{fig:b2xb1}
to constrain the frequency dependence of the observed signal.
If the signal at 150~GHz were due to synchrotron
we would expect the frequency cross spectrum
to be much larger in amplitude than the \biceptwo\ auto spectrum.
Conversely, if the 150~GHz power were due to polarized
dust emission we would not expect to see a significant correlation
with the 100~GHz sky pattern.

Pursuing this formally, we use simulations of
both experiments observing a common sky to construct a combined
likelihood function for band powers 1--5 of the \biceptwo\ auto,
\bicepone$_{100}$ auto, and their cross spectrum
using the Hamimeche-Lewis~\cite{hamimeche08} approximation (HL); see B14 for implementation details. 
As with all likelihood analyses we report, 
this procedure fully accounts for sample variance.
We use this likelihood function to fit a six-parameter model parametrized by
five 150~GHz band power amplitudes and a single common spectral index, $\beta$.
We consider two cases, in which the model accounts for (1) the total
$BB$ signal or (2) only the excess over lensed \lcdm, and
we take the spectral index to be the power
law exponent of this signal's antenna temperature as a function of
frequency. We marginalize this six-parameter model over the
band powers to obtain a one-parameter likelihood function over the
spectral index.

\begin{figure}
\begin{center}
\resizebox{0.7\columnwidth}{!}{\includegraphics{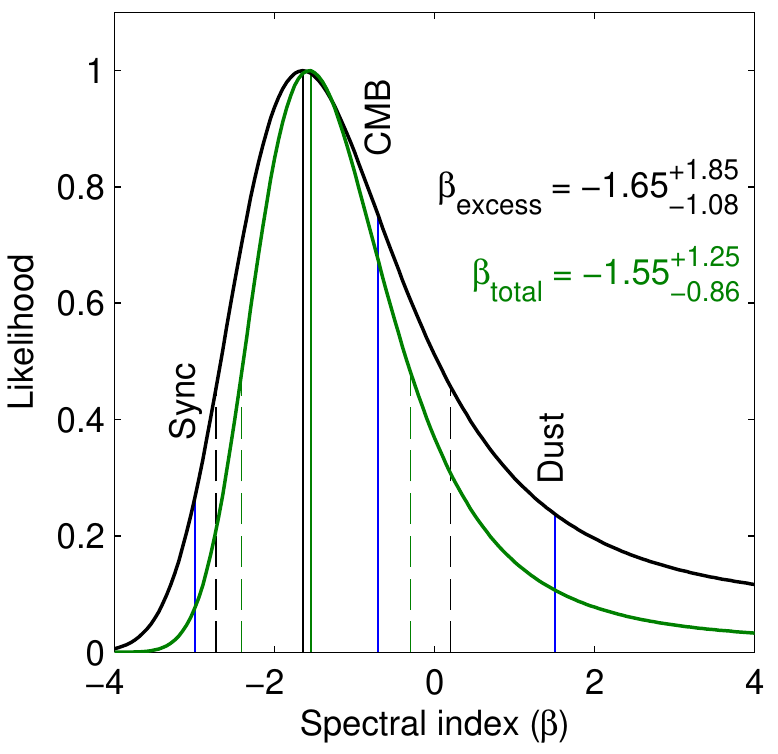}}
\end{center}
\caption{The constraint on the spectral index of the $BB$ total signal (green)
and excess signal over lensed \lcdm\ (black),
based on joint consideration of the \biceptwo\ auto, \bicepone$_{100}$ auto, and
\biceptwo$\times$\bicepone$_{100}$ cross spectra. The curve shows the
marginalized likelihood as a function of assumed spectral index. The
vertical solid and dashed lines indicate the maximum likelihoods and
the $\pm1\sigma$ intervals. The blue vertical lines indicate the
equivalent spectral indices under these conventions for the CMB,
synchrotron, and dust.
The observed signal is consistent with a CMB spectrum, while synchrotron
and dust are both disfavored.} 
\label{fig:specindcons}
\end{figure}

Figure~\ref{fig:specindcons} shows the resulting estimates of the
spectral index, with approximate 1$\sigma$ uncertainty ranges.
We evaluate the consistency with specific values of $\beta$ using a likelihood ratio
test.
Both the total and the excess observed $BB$ signal are 
consistent with the spectrum of the CMB 
($\beta=-0.7$ for these bands and conventions).
The spectrum of the excess $BB$ signal has a CMB-to-peak
likelihood ratio of $L=0.75$.
Following Wilks~\cite{wilks1938} we take $\chi^2 \approx -2 \log{L}$
and evaluate the probability to exceed this value of $\chi^2$
(for a single degree of freedom). 
A synchrotron spectrum with $\beta=-3.0$ is disfavored for
the excess $BB$ ($L=0.26$, PTE 0.10, $1.6\sigma$);
although the \biceptwo$\times$\wmap$_K$ spectrum offers a much stronger
constraint.
The preferred whole-sky dust spectrum from \planck~\cite{planckXI}, 
which corresponds under these conventions to 
$\beta\approx+1.5$, is also disfavored as an explanation for
the excess $BB$ ($L=0.24$, PTE 0.09, $1.7\sigma$).
We have also conducted a series of simulations applying this
procedure to simulated data sets with CMB and dust spectral indices.
These simulations indicate that the observed likelihood ratios are
typical of a CMB spectral index but atypical of dust
[For the dust simulations we
simulate power spectra for our sky patch using the HL likelihood
function, assuming the observed \biceptwo\ power spectrum at 150~GHz
and extrapolating to 100~GHz using a spectral index of $+1.5$ for
the excess above lensing. 
For each simulation we compute this likelihood function and calculate the likelihood
ratio of $L(1.5)/L(\textrm{CMB})$.
In 45 of 500 such simulations we find a likelihood ratio
smaller than that in our actual data.]

In the analysis above, the 100~GHz auto spectrum contributes
little statistical weight, so what is being constrained is effectively
the spectral index of the component of the 100~GHz sky pattern
which correlates with the 150~GHz pattern.
A mixture of synchrotron and dust, summing
to the level of the observed $BB$ excess,
could in principle be constructed to achieve 
any intermediate effective spectral index.
Spatial correlation between the two patterns is an additional
potential degree of freedom.
Considering a scenario with no such correlation and
nominal dust and synchrotron spectral
indices ($\beta_{sync}=-3.0$), reproducing the maximum likelihood effective 
$\beta = -1.65$ (see Fig.~\ref{fig:specindcons}) 
would require a nearly equal
mix of dust and synchrotron $BB$ power at 150~GHz.
In this scenario, the synchrotron contribution in
the \biceptwo\ auto spectrum would be $r_{sync,150}\approx0.10$. 
However, the corresponding constraint from the \biceptwo$\times$\wmap$_K$
cross spectrum (Sec.~\ref{sec:sync}, $r_{sync,150}=0.0014 \pm 0.0071$)
rules this scenario out at $13.5\sigma$. Calculating
the \bicepone$_{100}\times$\wmap$_K$ cross spectrum
yields a similar but slightly weaker constraint:
for $\beta_{sync}=-3.0$, $r_{sync,150}=-0.0005 \pm 0.0076$, 
disfavoring this scenario at $12.6\sigma$.

In a scenario with 100\% correlation between synchrotron and dust,
an effective index $\beta = -1.65$ can be produced with
a lower synchrotron contribution,
but the assumption of dust correlation adds to this scenario's
predicted level for \biceptwo$\times$\wmap$_K$, 
so that the actual measured cross spectrum also disfavors this scenario
at $> 13\sigma$.
More generally, scenarios which mix dust with 
synchrotron ($\beta_{sync}=-3.0$) with any assumed degree of 
correlation from $(0 - 100)\%$, in ratios needed to produce
an effective $\beta < 0.2$,
are disfavored by the \biceptwo$\times$\wmap$_K$ 
cross spectrum constraint at $> 3 \sigma$.  
Scenarios which would approximate a CMB-like index ($\beta=-0.7$)
with a mixture of dust and synchrotron are therefore unlikely.

\subsection{Additional cross spectra}

\begin{figure}
\begin{center}
\resizebox{\columnwidth}{!}{\includegraphics{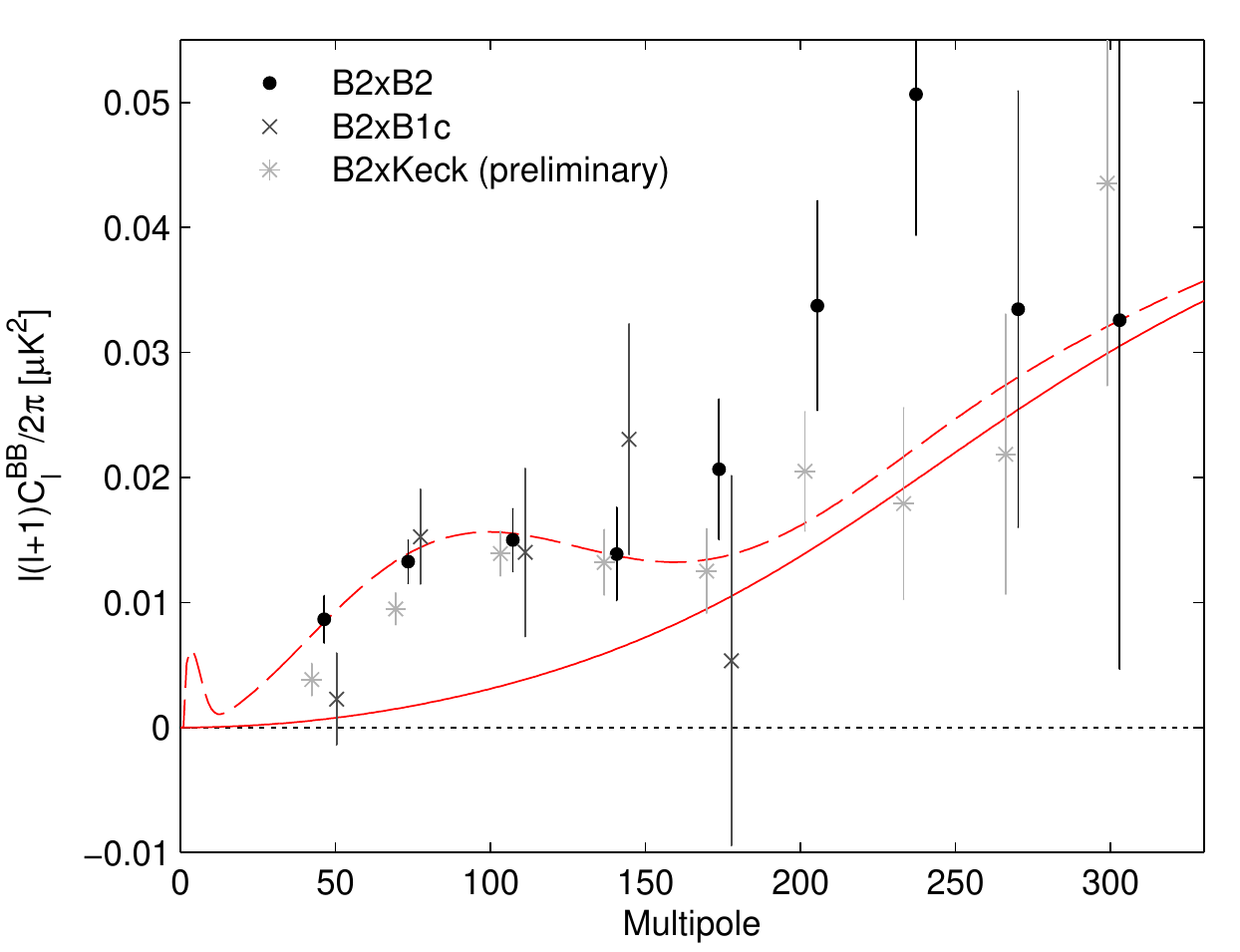}}
\end{center}
\caption{Comparison of the \biceptwo\ $BB$ auto spectrum and
cross spectra taken between \biceptwo\ and \bicepone\ combined,
and \biceptwo\ and \keckarray\ preliminary.
The error bars are the standard deviations of the
lensed-\lcdm+noise simulations and hence contain no sample variance on
tensors.
(For clarity the cross spectrum points are offset horizontally
and the \biceptwo$\times$\bicepone\ points are omitted at
$\ell>200$.)}
\label{fig:speccomp}
\end{figure}

Having seen that the \biceptwo\ auto spectrum is
compatible with both the \biceptwo$\times$\bicepone$_{100}$
and the \biceptwo$\times$\bicepone$_{150}$ cross spectra 
we proceed to combine the latter.
(We combine using weights which minimize the variance
of the lensed-\lcdm+noise simulations as described in B14.)
Figure~\ref{fig:speccomp} compares the result
to the \biceptwo\ auto spectrum from Fig.~\ref{fig:powspecres}.
Taking the difference of these spectra
and comparing to the differences of the lensed-\lcdm+noise+$r=0.2$
simulations the bandpower 1--5 $\chi^2$ and $\chi$
PTEs are mid-range indicating compatibility.

Comparing the \biceptwo$\times$\bicepone$_\mathrm{comb}$
spectrum to the lensed-\lcdm\ expectation the $\chi^2$
and $\chi$ values have PTE of $0.005$
and $0.002$, respectively, corresponding to
$\approx 3\sigma$ evidence of excess power.
The compatibility of the \biceptwo\ auto and
\biceptwo$\times$\bicepone$_\mathrm{comb}$ cross spectra
combined with the detection of excess power in the cross
spectra provides yet more evidence against a systematic
origin of the nominal signal given the significant
differences in focal plane technology and beam imperfections.

The successor experiment to \biceptwo\ is the \keckarray\ which
consists of five \biceptwo-like receivers \cite{sheehy10}.
The \keckarray\ data analysis is not yet complete and will be the
subject of future publications.
However, as an additional systematics check we show in
Fig.~\ref{fig:speccomp} a cross spectrum
between \biceptwo\ and preliminary \keckarray\ 150~GHz maps
from the 2012 and 2013 seasons.
This cross spectrum also shows obvious 
excess $BB$ power at low $\ell$.

\section{Cosmological Parameter Constraints}
\label{sec:parcons}

We have shown that our observed \bmode\ spectrum
(i) is not explained by known systematics (jackknifes, beam-map
simulations, other systematics studies, and cross spectra 
with \bicepone$_{150}$),
and (ii) domination by foregrounds is disfavored
(dust model projections, dust model cross correlations,
synchrotron constraints,
and spectral index constraints from cross spectra with \bicepone$_{100}$).
In this section we do some basic fitting of cosmological
parameters while noting again that all the band powers and ancillary
data are available for download so that others may conduct fuller
studies.

\subsection{Lensed-\lcdm\ + tensors}

In Fig.~\ref{fig:powspecres} we see a substantial excess
of $BB$ power in the region where an inflationary
gravitational wave (IGW) signal would be expected to peak.
We therefore proceed to find the most likely value of the
tensor-to-scalar ratio $r$ using the ``direct likelihood''
method introduced in B14.
We first form additional sets of simulations for many
values of $r$ by combining the lensed-\lcdm\ and scaled
$r=0.2$ simulations.
(Hence we assume always $n_t=0$ making the value of $r$ independent of
the tensor pivot scale.)
We then combine the band powers of these and the real band powers
with $s/n$ weighting where $s$ is the IGW spectrum for a small value
of $r$ and $n$ is the variance of the lensed-\lcdm+noise simulations.
Arranging the simulation pdf values as rows we can then
read off the likelihood curve for $r$ as the columns at the
observed combined band-power value.

\begin{figure*}
\begin{center}
\resizebox{\textwidth}{!}{\includegraphics{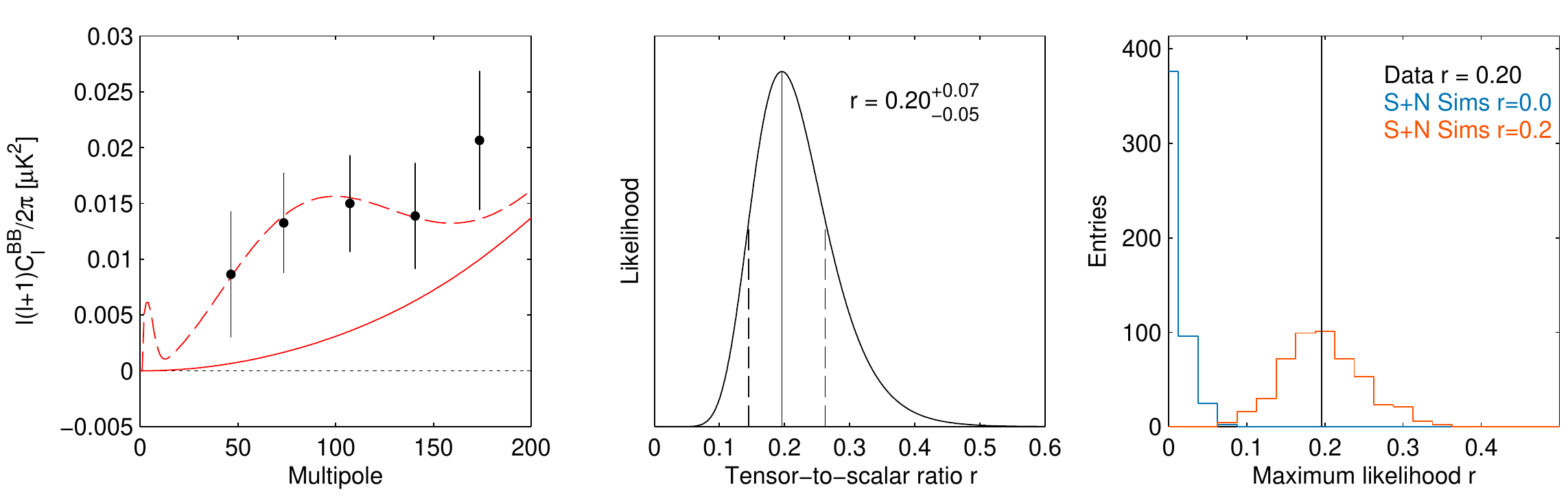}}
\end{center}
\caption{Left: The \biceptwo\ band powers plotted with the
maximum likelihood lensed-\lcdm+$r=0.20$ model.
The uncertainties are taken from that model and hence include
sample variance on the $r$ contribution.
Middle:
The constraint on the tensor-to-scalar ratio $r$.
The maximum likelihood and $\pm1\sigma$
interval is $r=0.20^{+0.07}_{-0.05}$, as indicated by the vertical lines.
Right:
Histograms of the maximum likelihood values of $r$ derived from
lensed-\lcdm+noise simulations with $r=0$ (blue) and adding $r=0.2$ (red).
The maximum likelihood value of $r$ for
the real data is shown by the vertical line.}
\label{fig:rlim}
\end{figure*}

The result of this process is shown in Fig.~\ref{fig:rlim}.
Defining the confidence interval as the equal likelihood
contour which contains 68\% of the total likelihood
we find $r=0.20^{+0.07}_{-0.05}$.
This uncertainty is driven by the sample variance in our
patch of sky, and the likelihood falls off very steeply towards
$r=0$.
The likelihood ratio between $r=0$ and the maximum is
$2.9\times10^{-11}$ equivalent to a PTE of $3.3\times10^{-12}$
or $7.0\sigma$.
The numbers quoted above are for bins 1--5 although due
to the weighting step they are highly insensitive to the inclusion
of the higher band powers.
(Absolute calibration and beam uncertainty are included
in these calculations but have a negligible effect.)

Evaluating our simple $\chi^2$ statistic between
band powers 1--5 and the lensed-\lcdm+noise+$r=0.2$
simulations yields a value of $1.1$, which for 4 degrees of freedom has
a PTE of $0.90$.
Using all nine band powers $\chi^2$ is 8.4, which for 8 degrees of
freedom has a PTE of $0.40$.
The model is therefore a perfectly acceptable fit to the data.

\begin{figure}
\begin{center}
\resizebox{0.7\columnwidth}{!}{\includegraphics{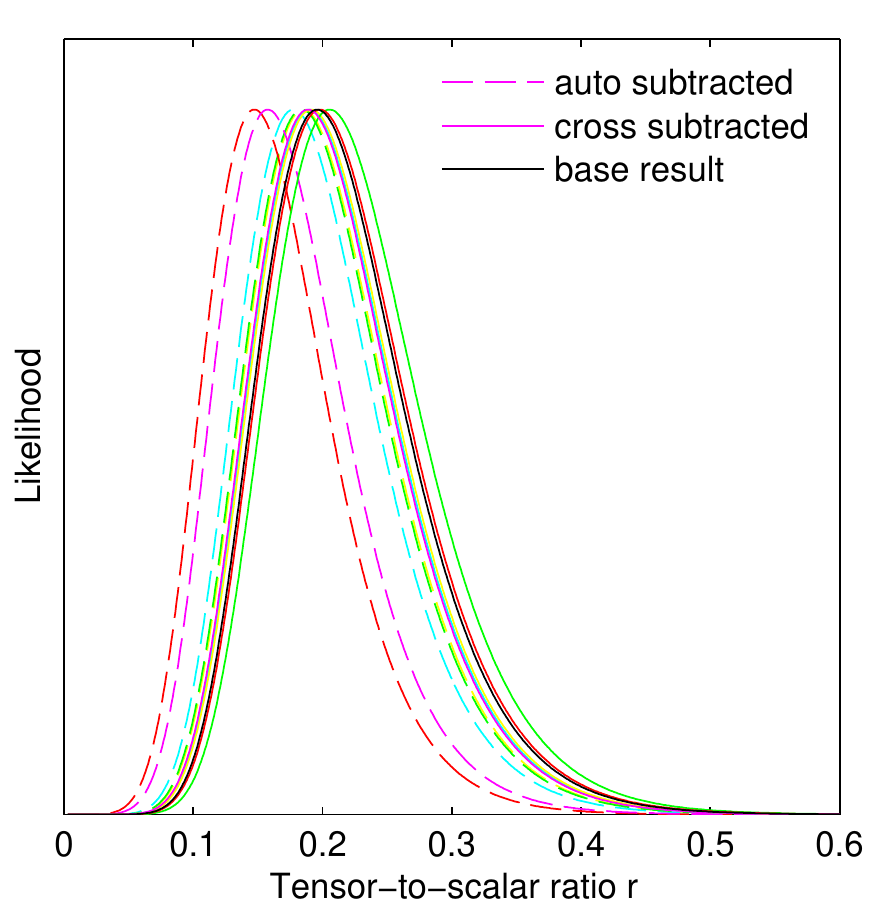}}
\end{center}
\caption{Modified constraints on the tensor-to-scalar ratio $r$ when
subtracting each of the foreground
models shown in Fig.~\ref{fig:foregroundproj} from the
\biceptwo\ $BB$ band powers.
The line styles and colors match Fig.~\ref{fig:foregroundproj}
with dashed for auto spectra and solid for cross spectra.
The probability that each of these models reflects
reality is hard to assess---see the text for discussion.
}
\label{fig:rcons_fgdebias}
\end{figure}

In Fig.~\ref{fig:rcons_fgdebias} we recompute the $r$ constraint
subtracting each of the dust models shown in
Fig.~\ref{fig:foregroundproj}.
For the auto spectra the range of maximum likelihood $r$
values is 0.15--0.19, while for the cross it is
0.19--0.21 (random fluctuations in the cross can cause
shifts up as well as down).
The probability that each of these models reflects
reality is hard to assess.
To explain the entire excess $BB$ signal with dust
requires increasing the power predicted by the auto spectra of the
various models by factors ranging from $\sim(5-10)\times$.
For example, in the context of the DDM1 model the preferred value of $r$ varies
as $r \sim 0.20 - 13 p^2$, so that increasing this model's assumed
uniform polarization fraction from $p = 5\%$ to
$p \sim 13\%$ would explain the full excess under this model.

The dust foreground is expected to have a
power law spectrum which slopes modestly down $\propto \ell^{\sim -0.6}$
in the usual $\ell(\ell+1)C_\ell/2\pi$ units~\cite{dunkley08}---although 
how this might fluctuate from small field to small
field at high Galactic latitude has not been investigated.
We note that the $s/n$ band-power weighting scheme described
above weights the first bin highly,
and it is here that the foreground models equal
the largest fraction of the observed signal.
Therefore if we were
to exclude the first band power the difference between the unsubtracted
and foreground subtracted model lines in Fig.~\ref{fig:rcons_fgdebias}
would be smaller;
i.e., while dust may contribute significantly
to our first band power it seems less able to explain band powers two
through five.
Reevaluating the base $r$ constraint using band powers 2--5
yields $r=0.19^{+0.07}_{-0.05}$ with $r=0$ ruled out at $6.4\sigma$.

Computing an $r$ constraint using the \biceptwo$\times$\bicepone$_\mathrm{comb}$
cross spectrum shown in Fig.~\ref{fig:speccomp} yields $r=0.19^{+0.11}_{-0.08}$.
The likelihood ratio between $r=0$ and the maximum is $2.0\times10^{-3}$
equivalent to a PTE of $4.2\times10^{-4}$ or $3.5\sigma$.

\subsection{Scaled-lensing + tensors}

Lensing deflections of the CMB photons as they travel from last
scattering remap the patterns slightly.
In temperature this leads to a slight smoothing of the
acoustic peaks, while in polarization a small $B$ mode is
introduced with a spectrum similar to a smoothed version
of the $EE$ spectrum a factor $\sim100$ lower in power.
Using their own and other data \planck~\cite{planckXVI} quote
a limit on the amplitude of the lensing effect versus
the \lcdm\ expectation of $A_L=0.99\pm0.05$.

\begin{figure}
\begin{center}
\resizebox{0.7\columnwidth}{!}{\includegraphics{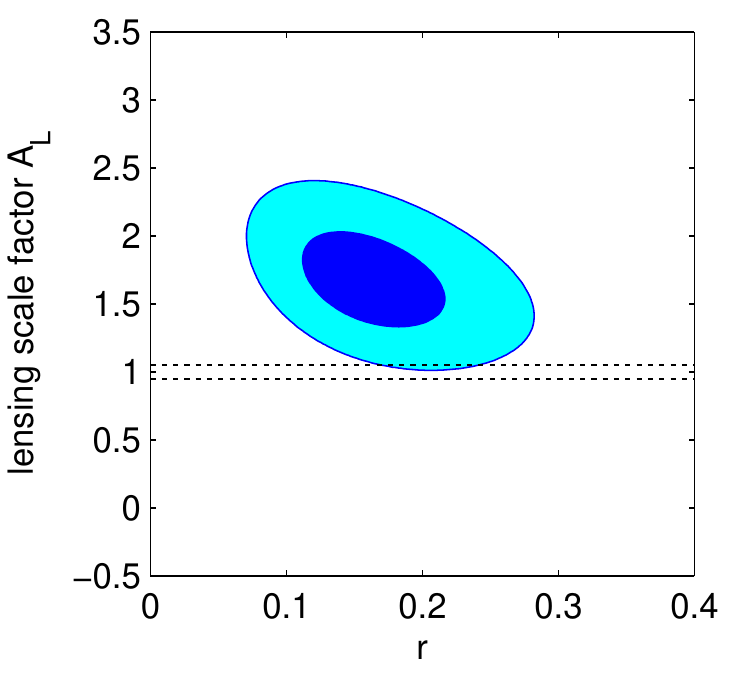}}
\end{center}
\caption{Joint constraints on the tensor-to-scalar
ratio $r$ and the lensing scale factor $A_L$ using the
\biceptwo\ $BB$ band powers 1--9.
One and two $\sigma$ contours are shown.
The horizontal dotted lines show the $1\sigma$ constraint from 
\planck~\cite{planckXVI}.
The \bmode\ lensing signal is detected at $5.5\sigma$,
with an amplitude $\sim 2\sigma$ higher than the expected value.}
\label{fig:rvsl}
\end{figure}

Figure~\ref{fig:rvsl} shows a joint constraint on the tensor-to-scalar
ratio $r$ and the lensing scale factor $A_L$ using our
$BB$ band powers 1--9.
As expected there is a weak anticorrelation---one can partially
explain the low $\ell$ excess by scaling up the lensing signal.
However, the constraint is mostly driven by band powers six
through nine
where the IGW signal is small.
The maximum likelihood scaling is $\approx 1.75$,
$\sim 2 \sigma$ from unity.
Marginalizing over $r$ the likelihood ratio between peak and
zero is $3\times10^{-7}$, equivalent to a PTE of $4.7\times10^{-8}$ or a $5.5\sigma$
detection of lensing in the \biceptwo\ $BB$ auto spectrum.
We note again that the high values of band powers
six and seven are not present in the preliminary cross
spectra against \keckarray\ shown in Fig.~\ref{fig:speccomp}.

\subsection{Compatibility with temperature data}

If present at last scattering, tensor modes will add power
to all spectra including $TT$.
For an $r$ value of 0.2 the contribution to $TT$
at the largest angular scales ($\ell<10$) would be $\approx 10\%$
of the level measured by \wmap\ and \planck.
The theoretical \lcdm\ power level expected at these scales is
dependent on several cosmological parameters including
the spectral index of the initial scalar perturbations $n_s$
and the optical depth to the last scattering surface $\tau$.
However, by combining temperature data taken over a
wide range of angular scales indirect limits on $r$ have been set.
A combination of \wmap+\spt\ data~\cite{story13} yields
$r<0.18$ (95\% confidence) tightening to $r<0.11$ when also including
measurements of the Hubble constant and baryon acoustic oscillations
(BAO).
More recently \planck~\cite{planckXVI} quote $r<0.11$ using a combination
of \planck, \spt\ and \act\ temperature data, plus \wmap\ polarization
(to constrain $\tau$).

These limits appear to be in moderately strong tension with
interpretation of our \bmode\ measurements as primarily due to
tensors.
One possibility is a larger than anticipated contribution
from polarized dust, but as our present data disfavor this
one can ask what additional extensions to the standard model might
resolve the situation.

\begin{figure}
\begin{center}
\resizebox{\columnwidth}{!}{\includegraphics{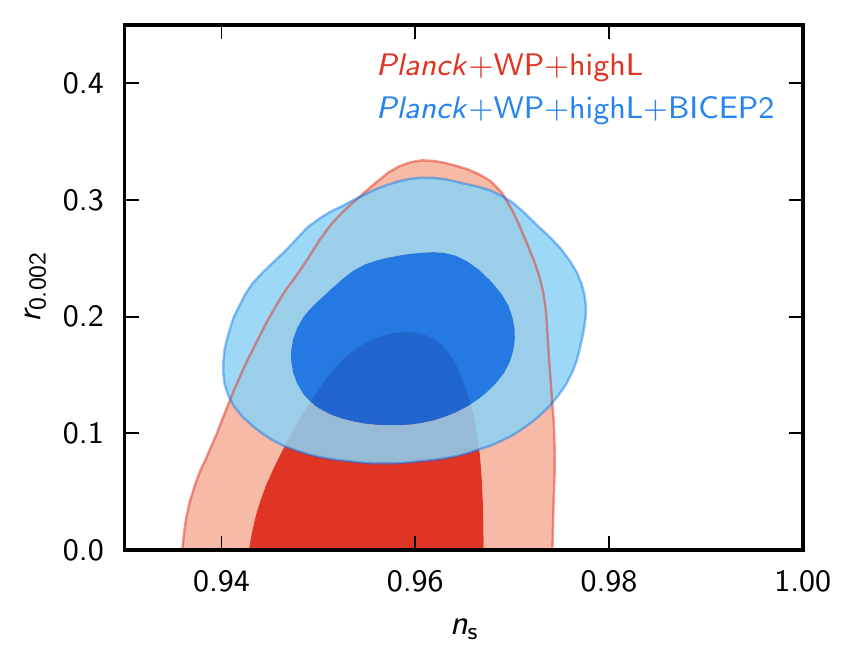}}
\end{center}
\caption{
Indirect constraints on $r$ from CMB temperature spectrum measurements
relax in the context of various model extensions.
Shown here is one example, following \planck\ XVI~\cite{planckXVI} Fig.~23,
where tensors and running of the scalar spectral index
are added to the base \lcdm\ model.
The contours show the resulting 68\% and 95\% confidence regions
for $r$ and the scalar spectral index $n_s$ when also allowing running.
The red contours are for the ``Planck+WP+highL'' data combination,
which for this model extension gives a 95\% bound $r<0.26$~\citep{planckXVI}.
The blue contours add the \biceptwo\ constraint on $r$ shown in the
center panel of Fig.~\ref{fig:rlim}.
See the text for further details.}
\label{fig:running}
\end{figure}

One obvious modification is to allow the initial scalar
perturbation spectrum to depart from the simple power
law form which is assumed in the base \lcdm\ model.
A standard way in which this is done is
by introducing a ``running'' parameter $dn_s/d\ln{k}$.
In \planck\ XVI~\cite{planckXVI} the constraint relaxes to $r<0.26$
(95\% confidence) when running is allowed with
$dn_s/d\ln{k} =-0.022\pm0.010$ (68\%) (for the
\planck+WP+highL data combination).
In Fig.~\ref{fig:running} we show the constraint contours
when allowing running as taken from Fig.~23 of \cite{planckXVI},
and how these change when the \biceptwo\ data are added.
The red contours on the plot are simply the Monte
Carlo Markov chains (MCMC)~\cite{gamerman06, lewis02} provided
with the \planck\ data release~\footnote{
As downloaded from \url{http://www.sciops.esa.int/wikiSI/planckpla}
section ``Cosmological Parameters.''} (and are thus
identical to those shown in that \planck\ paper).
We then apply importance sampling~\cite{hastings70} to these chains
using our $r$ likelihood as shown in Fig.~\ref{fig:rlim}
to derive the blue contours, 
for which the running parameter constraint shifts to  
$dn_s/d\ln{k} =-0.028\pm0.009$ (68\%).

The point of Fig.~\ref{fig:running} is not to endorse
running as the correct explanation of the observed deficit of
low $\ell$ $TT$ power.
It is simply to illustrate one example of a 
simple model extension beyond standard \lcdm+tensors
which can resolve the apparent tension between previous
$TT$ measurements and the direct evidence for tensors provided
by our \bmode\ measurements---probably there are others.
Of course, one might also speculate that
the tension could be reduced within the standard
\lcdm+tensors model, for example if $\tau$ or other
parameters were allowed to shift.
We anticipate a broad range of possibilities will be explored.

\section{Conclusions}
\label{sec:conc}

We have described the observations, data reduction, simulation,
and power spectrum analysis of all three seasons
of data taken by the \biceptwo\ experiment.
The polarization maps presented here are the deepest
ever made at degree angular scales having noise level
of 87~nK-deg in $Q$ and $U$ over an effective area
of 380 square deg.

To fully exploit this unprecedented sensitivity we have
expanded our analysis pipeline in several ways.
We have added an additional filtering of the time stream
using a template temperature map (from \planck) to
render the results insensitive to temperature to polarization
leakage caused by leading order beam systematics.
In addition we have implemented a map purification step that
eliminates ambiguous modes prior to \bmode\ estimation.
These deprojection and purification steps
are both straightforward extensions of the kinds of
linear filtering operations that are now common in CMB 
data analysis.

The power spectrum results are consistent
with lensed-\lcdm\ with one striking exception: the detection of
a large excess in the $BB$ spectrum in the
$\ell$ range where an inflationary gravitational wave signal
is expected to peak.
This excess represents a $5.2\sigma$ excursion from the
base lensed-\lcdm\ model.
We have conducted a wide selection of jackknife tests
which indicate that the \bmode\ signal is common on the sky
in all data subsets.
These tests offer strong empirical evidence against a
systematic origin for the signal.

In addition, we have conducted extensive simulations using
high fidelity per channel beam maps.
These confirm our understanding of the beam effects, and
that after deprojection of the two leading order modes,
the residual is far below the level of the signal which 
we observe.

Having demonstrated that the signal is real and ``on the sky''
we proceeded to investigate if it may be due to foreground
contamination.
Polarized synchrotron emission from our galaxy is estimated
to be negligible using low frequency polarized maps from \wmap.
For polarized dust emission public maps are not yet available.
We therefore investigate a number of 
commonly used models and
one which uses information which is currently
officially available from \planck.
At default parameter values
these models predict auto spectrum power well below our observed level.
However, these models are not yet well constrained by external
public data, which cannot empirically exclude dust emission 
bright enough to explain the entire excess signal.
In the context of the DDM1 model,
explaining the entire excess signal would
require increasing the predicted dust power spectrum
by $6 \times$, for example by increasing the assumed
uniform polarization fraction in our field from 
5\% (a typical value) to $\sim 13\%$. 
None of these models show significant cross-correlation
with our maps (although this may be interpreted simply
as due to limitations of the models).

Taking cross spectra against 100~GHz maps from \bicepone\ we
find significant correlation and set a constraint
on the spectral index of the \bmode\ excess consistent with
CMB and disfavoring dust by $1.7\sigma$.
The fact that the \bicepone\ and \keckarray\ maps cross correlate
with \biceptwo\ is powerful further evidence against systematics.

An economical interpretation
of the \bmode\ signal which we have detected is that it
is largely due to tensor modes---the IGW template is an excellent
fit to the observed excess.
We therefore proceed to set a constraint on the tensor-to-scalar ratio
and find $r=0.20^{+0.07}_{-0.05}$ with $r=0$ ruled out
at a significance of $7.0\sigma$, with no foreground subtraction.
Multiple lines of evidence suggest that the contribution of
foregrounds (which will lower the favored value of $r$) is
subdominant:
(i) direct projection of the available foreground models
using typical model assumptions, 
(ii) lack of strong cross-correlation of those models against
the observed sky pattern (Fig.~\ref{fig:foregroundproj}),
(iii) the frequency spectral index of the signal as constrained
using \bicepone\ data at 100~GHz (Fig.~\ref{fig:specindcons}), and
(iv) the power spectral form of the signal and its apparent spatial isotropy
(Figs.~\ref{fig:eb_maps} and~\ref{fig:rlim}).

Subtracting the various dust models at their default parameter values
and re-deriving the $r$ constraint still results in high significance of detection.
As discussed above, one possibility that cannot be ruled out
is a larger than anticipated contribution from polarized dust. 
Given the present evidence disfavoring this,
these high values of $r$ are in apparent tension with
previous indirect limits based on temperature measurements and we have discussed some
possible resolutions including modifications
of the initial scalar perturbation spectrum such as running.
However, we emphasize that we do not claim to know what the resolution
is, if one is in fact necessary.

\begin{figure}
\resizebox{\columnwidth}{!}{\includegraphics{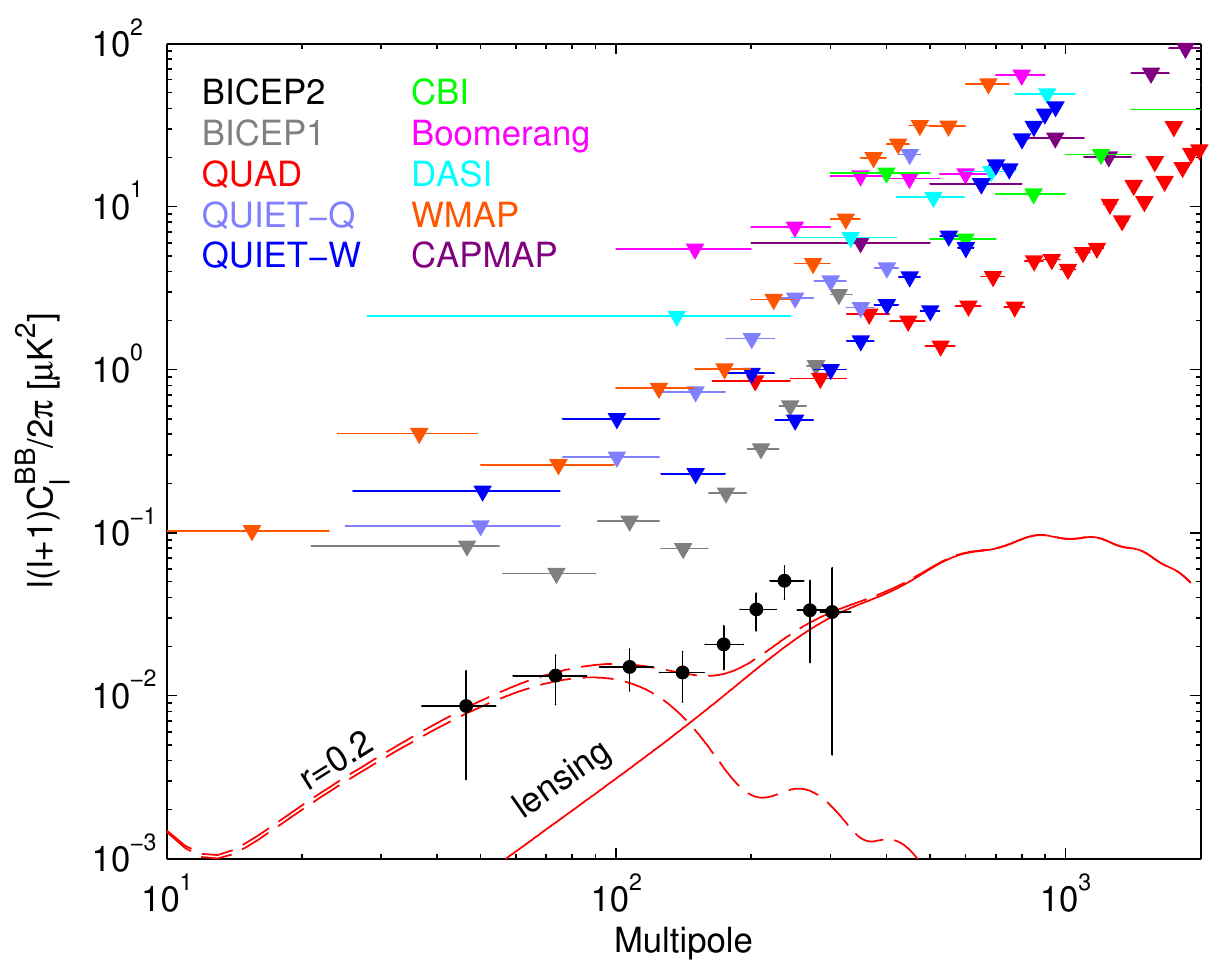}}
\caption{\biceptwo\ $BB$ auto spectra and 95\% upper limits
from several previous experiments~\cite{leitch05, montroy06,sievers07, bischoff08, brown09, quiet11, quiet12, bennett13, barkats14}.
The curves show the theory expectations
for $r=0.2$ and lensed-\lcdm.
The \biceptwo\ uncertainties include sample variance on an $r=0.2$ contribution.
}
\label{fig:b2_and_previous_limits}
\end{figure}

Figure~\ref{fig:b2_and_previous_limits} shows the \biceptwo\ results
compared to previous upper limits.
We have pushed into a new regime of sensitivity, and
the high-confidence detection of \bmode\ polarization 
at degree angular scales brings us to an exciting juncture.
If the origin is in tensors, as favored by 
the evidence presented above,
it heralds a new era of \bmode\ cosmology.
However, if these $B$ modes represent evidence of a high-dust foreground,
it reveals the scale of the challenges that lie ahead.

\begin{acknowledgments}

\biceptwo\ was supported by the U.S. National Science Foundation under 
Grants No.\ ANT-0742818 and ANT-1044978 (Caltech and Harvard) 
and ANT-0742592 and ANT-1110087 (Chicago and Minnesota).  
The development of antenna-coupled detector technology was supported 
by the JPL Research and Technology Development Fund and Grants No.\
06-ARPA206-0040 and 10-SAT10-0017
from the NASA APRA and SAT programs.  
The development and testing of focal planes were supported 
by the Gordon and Betty Moore Foundation at Caltech.  
Readout electronics were supported by a Canada Foundation
for Innovation grant to UBC.
The receiver development was supported in part 
by a grant from the W.M.\ Keck Foundation.  
The computations in this paper were run on the Odyssey cluster
supported by the FAS Science Division Research Computing Group at
Harvard University.
The analysis effort at Stanford and SLAC is partially supported by 
the U.S. Department of Energy Office of Science.
Tireless administrative support was provided by Irene Coyle
and Kathy Deniston.
We thank the staff of the U.S. Antarctic Program and in particular 
the South Pole Station without whose help this research would not have been possible.
We thank all those who have contributed past efforts to the \bicep--\keckarray\
series of experiments, including the \bicepone\ and \keckarray\ teams.
We thank all those in the astrophysics community who have contributed
feedback on the public preprint of this paper, and particularly two
anonymous referees for their detailed and constructive recommendations. 
This work would not have been possible without the late Andrew Lange,
whom we sorely miss.

\end{acknowledgments}

\subsubsection*{\textbf{Note added}}

Since we submitted this paper new information on polarized dust emission
has become available from the \planck\ experiment in a series of
papers~\cite{planckiXIX,planckiXX,planckiXXI,planckiXXII}.
While these confirm that the modal polarization fraction of dust is
$\sim 4$\%, there is a long tail to fractions as high as 20\%
(see Fig. 7 of~\cite{planckiXIX}).
There is also a trend to higher polarization fractions in
regions of lower total dust emission
[see Fig.~18 of \cite{planckiXIX} noting that the \biceptwo\ field
has a column density of $\sim(1-2)\times 10^{20}$~H~cm$^{-2}$].
We note that these papers 
restrict their analysis to regions of the sky where
``systematic uncertainties are small, and where the dust signal dominates
total emission,'' and that this excludes 21\% of the sky that includes
the \biceptwo\ region.
Thus while these papers do not offer definitive information on
the level of dust contamination in our field, they do suggest
that it may well be higher than any of the models considered in
Sec.~\ref{sec:foregrounds}.

In addition there has been extensive discussion of our preprint
in the cosmology community.
Two preprints~\cite{fuskeland14,flauger14} look at polarized synchrotron
emission in our field and conclude that at 150~GHz it is very small, in broad agreement
with our analysis in Sec.~\ref{sec:foregrounds}.
Several preprints also examine the new information from \planck\
and raise the same concerns discussed above---that 
the polarized dust emission may be stronger
than any of the models discussed in Sec.~\ref{sec:foregrounds}~\cite{flauger14,mortonson14}.
Given these concerns these studies also reexamine our spectral index constraint
described in Sec.~\ref{sec:specindcons},
since this offers (weak) evidence that the signal is not dust dominated.
Both point out that our initial analysis gave the effective spectral
index of the whole signal---including the lensing component.
Figure~\ref{fig:specindcons} now shows an additional curve for the
excess over lensing only---the maximum likelihood value is nearly unchanged
while the evidence against dust is somewhat weakened.
However, Flauger \textit{et al.}~\cite{flauger14} also question whether
sample variance is properly included in our spectral index analysis,
and whether noise in foreground templates could systematically 
suppress our estimates of template cross correlation.
In fact, sample variance is naturally included in the HL-based formalism
on which our spectral analysis is based,
and the template cross spectra we report 
are not subject to bias from noise.

More data is clearly required to resolve the situation.
We note that cross-correlation of our maps with the \planck\
353~GHz maps will be more powerful than use of those maps
alone in our field.
Additional data are also expected from many other experiments, 
including \keckarray\ observations at 100~GHz in the 2014 season.

\bibliography{ms}

\end{document}

%% file: defs.tex
\def\act{{\sc Act}}
\def\bicep{BICEP}

\def\dasi{DASI}

\def\bicepone{BICEP1}
\def\biceptwo{BICEP2}

\def\keckarray{{\it Keck Array}}
\def\planck{{\it Planck}}
\def\quiet{QUIET}
\def\spider{SPIDER}

\def\QUAD{QUAD}
\def\cbi{CBI}
\def\capmap{CAPMAP}
\def\maxipol{MAXIPOL}

\def\ebex{EBEX}
\def\boom{BOOMERANG}
\def\wmap{WMAP}
\def\spt{SPT}

\def\polarbear{POLARBEAR}
\def\abs{ABS}
\def\actpol{ACTPOL}

\def\class{CLASS}
\def\piper{PIPER}

\def\camb{CAMB}

\def\synfast{{\sc synfast}}
\def\healpix{{\sc healp}ix}
\def\lenspix{{\sc lenspix}}
\def\master{MASTER}

\newcommand{\ukrts}{ $\mu\mathrm{K}_{\mathrm{\mbox{\tiny\sc cmb}}}\sqrt{\mathrm{s}}$}

\def\deg{^\circ}
\def\emode{$E$-mode}
\def\bmode{$B$-mode}

\def\lcdm{$\Lambda$CDM}

\def\sovast{Sov.\ Astron.}
\def\aap{Astron.\ Astrophys.}

\def\apl{Appl.\ Phys.\ Lett.}

\def\apj{Astrophys.\ J.}
\def\apjl{Astrophys.\ J.\ Lett.}
\def\apjs{Astrophys.\ J.\ Suppl.\ Ser.}

\def\jcap{J.\ Cosmol.\ Astropart.\ Phys.}

\def\mnras{Mon.\ Not.\ R.\ Astron.\ Soc.}

\def\prd{Phys.\ Rev.\ D}

\def\prl{Phys.\ Rev.\ Lett.}

%% file: ms.bbl
\begin{thebibliography}{113}
\bibitem [{\citenamefont {{\textsc{Bicep2} Collaboration
  III}}(shed)}]{b2systpap14}%
  \BibitemOpen
  \bibfield  {author} {\bibinfo {author} {\bibnamefont {{\textsc{Bicep2}
  Collaboration III}}},\ }\href@noop {} {\  (\bibinfo {year} {to be
  published})}\BibitemShut {NoStop}%
\end{thebibliography}
